%% file: Main-Revised-Clean.tex
\documentclass[aps,reprint,
superscriptaddress,nofootinbib,longbibliography,noeprint
]{revtex4-2}
\usepackage{physics}
\usepackage{graphicx}
\usepackage{dcolumn}
\usepackage{amsmath}
\usepackage{amssymb}
\usepackage{graphicx}
\usepackage{wrapfig}
\usepackage{xcolor}
\definecolor{dodgerblue}{HTML}{1E90FF}
\definecolor{lightdodgerblue}{HTML}{4dbff7}
\usepackage[colorlinks=true,citecolor=dodgerblue,linkcolor=dodgerblue,urlcolor=dodgerblue,pdftitle={RSRC}]{hyperref}
\usepackage{float}
\usepackage{yfonts}
\usepackage{hhline}
\usepackage{bbm}
\usepackage{bm}
\renewcommand{\vec}{\bm}

\def\equationautorefname~#1\null{Eq. (#1)\null}

\newcommand{\pg}[1]{\hat{\mathcal{P}}^{#1}}
\usepackage[makeroom]{cancel}
\newcommand{\micro}[1]{\ensuremath{\mathcal{N}^{(#1)}}}
\newcommand{\depol}{\mathcal{N}_p}
\newcommand{\Ic}{\ensuremath{I_c(R\rangle Q)}}
\newcommand{\appref}[1]{\hyperref[#1]{App.~\ref*{#1}}}

\newcommand{\vn}{\text{vN}}
\newcommand{\haar}{\mathbb{E}_U}
\newcommand{\kket}[1]{| #1 \rangle \! \rangle}
\newcommand{\bbra}[1]{\langle \! \langle #1 |}
\newcommand{\Wg}[2]{\text{Wg}^{(#1)}_{#2}}
\newcommand{\ztr}{\text{tr}} 
\newcommand{\zlog}{\text{log}} 

\usepackage[caption=false]{subfig}

\usepackage[normalem]{ulem}
\makeatletter 
    
\renewcommand\onecolumngrid{
\do@columngrid{one}{\@ne}%
\def\set@footnotewidth{\onecolumngrid}
\def\footnoterule{\kern-6pt\hrule width 1.5in\kern6pt}%
}

\renewcommand\twocolumngrid{
        \def\footnoterule{
        \dimen@\skip\footins\divide\dimen@\thr@@
        \kern-\dimen@\hrule width.5in\kern\dimen@}
        \do@columngrid{mlt}{\tw@}
}%

\makeatother    

\begin{document}
\newcommand{\lhead}[1]{\section{#1}}
\renewcommand{\grad}{\vec{\nabla}}
\renewcommand{\curl}{\vec{\nabla}\times}
\renewcommand{\divergence}{\vec{\nabla}\cdot}
\renewcommand{\l}{\left(}
\renewcommand{\r}{\right)}
\newcommand{\lb}{\left[}
\newcommand{\rb}{\right]}
\newcommand{\lcb}{\left\{ }
\newcommand{\rcb}{\right\} }
\newcommand{\comment}[1]{}
\newcommand{\lv}{\left|}
\newcommand{\rv}{\right|}
\newcommand{\mbf}[1]{\mathbf{\mathrel{#1}}}
\newcommand{\mc}[1]{\mathcal{\mathrel{#1}}}
\newcommand{\msc}[1]{\mathscr{\mathrel{#1}}}
\newcommand{\up}{\uparrow}
\newcommand{\down}{\downarrow}
\newcommand{\pI}{\mathbb{I}_{2}}
\newcommand{\pX}{\mathbf{X}}
\newcommand{\pY}{\mathbf{Y}}
\newcommand{\pZ}{\mathbf{Z}}
\newcommand{\npI}{\hat{\mathbb{I}}_{2}}
\newcommand{\npX}{\hat{\mathbf{X}}}
\newcommand{\npY}{\hat{\mathbf{Y}}}
\newcommand{\npZ}{\hat{\mathbf{Z}}}
\newcommand{\dbar}{\mathchar'26\mkern-12mu d}
\renewcommand{\log}[1]{\, {\rm Log} \left[ \mathrel{#1} \right] }
\newcommand{\titlemath}[1]{\texorpdfstring{$\mathrel{#1}$}{TEXT}}
\renewcommand{\=}[1]{$\mathrel{#1}$}
\renewcommand{\tr}[1]{{\rm Tr}\lb\mathrel{#1}\rb}
\newcommand{\ptr}[2]{{\rm Tr}_{#2}\lb\mathrel{#1}\rb}
\newcommand{\svn}[1]{S_{\rm vN}^{\rm \mathrel{#1}}}
\newcommand{\mbs}[1]{\boldsymbol{#1}}
\renewcommand{\mod}[1]{\, {\rm mod} \; #1}
\newcommand{\oft}{\l  t \r}
\newcommand{\lattice}{\mathscr{L}}

\newcommand{\gbra}[1]{\langle{#1}|}
\newcommand{\gket}[1]{|{#1}\rangle}
\newcommand{\glog}{\mathrm{log}}
\newcommand{\stab}{P}
\newcommand{\logic}{P'}

\def\beq{\begin{equation}}
\def\eeq{\end{equation}}
\def\bea{\begin{eqnarray}}
\def\eea{\end{eqnarray}}

\preprint{APS/123-QED}
\title{Spectral properties and coding transitions of Haar-random quantum codes}

\author{Grace~M.~Sommers}
\affiliation{Department of Physics, Princeton University, Princeton, New Jersey 08544, USA}
\author{J.~Alexander~Jacoby}
\affiliation{Department of Physics, Princeton University, Princeton, New Jersey 08544, USA}
\affiliation{Max Planck Institute for the Physics of Complex Systems, 01187 Dresden, Germany}
\author{Zack Weinstein}
\affiliation{Department of Physics, University of California, Berkeley, California 94720, USA}
\affiliation{Department of Physics and Institute for Quantum Information and Matter, California Institute of Technology, Pasadena, California 91125, USA}
\author{David~A.~Huse}
\affiliation{Department of Physics, Princeton University, Princeton, New Jersey 08544, USA}
\author{Sarang~Gopalakrishnan}
\affiliation{Department of Electrical and Computer Engineering, Princeton University, Princeton, New Jersey 08544, USA}


\begin{abstract}
    A quantum error-correcting code with a nonzero error threshold undergoes a mixed-state phase transition when the error rate reaches that threshold. We explore this phase transition for Haar-random quantum codes, in which the logical information is encoded in a random subspace of the physical Hilbert space. We focus on the spectrum of the encoded system density matrix as a function of the rate of uncorrelated, single-qudit errors. For low error rates, this spectrum consists of well-separated bands, representing errors of different weights. As the error rate increases, the bands for high-weight errors merge. The evolution of these bands with increasing error rate is well described by a simple analytic ansatz. Using this ansatz, as well as an explicit calculation, we show that the threshold for Haar-random quantum codes saturates the hashing bound, and thus coincides with that for random \emph{stabilizer} codes. For error rates that exceed the hashing bound, typical errors are uncorrectable, but postselected error correction remains possible until a much higher \emph{detection} threshold. Postselection can in principle be implemented by projecting onto subspaces corresponding to low-weight errors, which remain correctable past the hashing bound. 
\end{abstract}

\maketitle

\section{Introduction}

The threshold theorem for quantum error correcting codes is a landmark result in quantum information theory~\cite{shor1996fault,aharonov1997,kitaev1997,Knill1998,aharonov2008}. In recent work, the error correction threshold has drawn interest as an example of a \emph{mixed-state phase transition} \cite{Sang_24, CWvK_24,Ma2025}. Encoding a logical state in a finite-threshold error-correcting code, and subjecting the encoded state to local decoherence of strength $p$, defines a family of mixed states $\rho^{\l  p \r}$. When $p$ is below an error threshold $p_c$, the mixed state $\rho^{\l  p \r}$ retains logical information; past the threshold, logical information cannot be reliably recovered even with an optimal decoder \cite{Colmanarez2024,Lee2025coherent,Niwa2025,Colmanarez2025}. The distinction between these regimes is an intrinsic \textit{information-theoretic} property of the mixed states. In particular, conventional local correlation functions are insensitive to this transition: all the mixed states $\rho^{\l  p \r}$ are derived from the error-free state $\rho^{(p = 0)}$ by one round of single-qudit channels, which can only modify local correlation functions analytically as $p$ is varied through $p_c$, even in the large-system limit \cite{FanBao_24}.

A large body of recent work \cite{bao2023mixedstatetopologicalordererrorfield,PRXQuantum.4.030317,FanBao_24,Chen_2024, Lyons_24,Hauser2026} has focused on identifying changes in the entanglement structure of $\rho^{\l  p \r}$ across the error correction threshold, primarily focusing on topological codes such as the toric code \cite{Kitaev_2003, Dennis2002, Bombin_12}. For example, it has been argued that $\rho^{\l  p \r}$ can be written as an ensemble of short-range entangled states only if $p > p_c$~\cite{Chen_2024,Chen2024symm}); the conditional correlation length (Markov length) of $\rho^{\l  p \r}$ is thought to diverge at $p_c$ \cite{Sang2025markov,negari2025spacetimemarkovlengthdiagnostic}; and the von Neumann entropy of $\rho^{\l  p \r}$ is nonanalytic at $p_c$~\cite{Su_24,Lyons_24}. Moreover, it was pointed out in Refs.~\cite{FanBao_24,Su_24,Lyons_24}  that there are further transitions in the entanglement structure that occur \emph{past} the error correction threshold, associated with singularities in the R\'enyi entropies of $\rho^{\l  p \r}$. The information-theoretic implications (if any) of these further transitions have not yet been explained. 

\begin{figure}[b]
\centering
\includegraphics[width=0.7\linewidth]{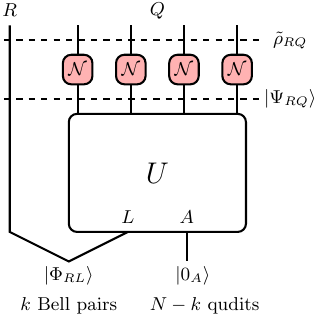}
\caption{Schematic circuit diagram of the Haar-random codes studied in this work. $k$ logical qudits are isometrically encoded in an $N$-qudit system via a Haar-random unitary $U$. Each system qudit $i$ is then subjected to a depolarizing channel $\mathcal{N}_{i,p}$ of strength $p$ [see Eq.~\eqref{eqn:dep_channel3-p}].
\label{fig:noisy-encoding}}
\end{figure}

In the present work we explore the nature of this mixed-state phase transition, not for highly structured models like the toric code, but for \emph{random} codes. A random code embeds the logical Hilbert space into a subspace of the (typically much larger) physical Hilbert space via a Haar-random isometry. The physical Hilbert space is taken to have a tensor-product structure, consisting of $N$ qudits (each of dimension $q$), and the noise model we consider is single-qudit depolarization acting on each physical qudit with a strength $p$ (defined more precisely below). The only structure in the model comes from the tensor product structure and the single-qudit nature of the noise.

The unstructured nature of these codes enables a novel semi-analytic treatment of the spectrum of the mixed state $\rho^{\l  p \r}$. For very small $p$, this spectrum consists of ``bands'' of eigenvalues that are well-separated on a logarithmic scale. To our knowledge, this band structure has gone unnoted in the literature until this work. The $w$th band consists of states that (to a good approximation) are of weight $w$, meaning that they can be created from the encoded logical state by applying errors on $w$ sites. Thus, for qudits there are $\l q^{2} -1 \r^{w} \binom{N}{w}$ states in the $w$th band and their eigenvalues scale as $\sim p^w$ at small $p$. Each error maps the logical subspace to approximately orthogonal subspaces; small-$w$ errors are typically correctable since these (relatively dilute) subspaces are not forced to overlap. 
Since for $N$ qudits there are $q^{2N}$ possible errors, and only $O(q^{N})$ mutually orthogonal subspaces (for $O(1)$ encoded logical qudits), not all errors can be corrected: instead, errors with weight $w > w_c \approx 0.189 N$ for qubits (the quantum hashing bound \cite{Bennett1996hashing}) do not map the logical subspace to dilute approximately orthogonal subspaces, and thus cannot be corrected. The band with the smallest eigenvalues is associated with these uncorrectable errors. 

The description above is parametrically accurate as $p\to 0$. As one increases $p$, the bands become less well separated, and inter-band mixing of error weights becomes important. We treat this band mixing perturbatively and find good agreement with numerics for a single logical qubit (\autoref{fig:canonical}). Crucially, for any sub-threshold $p<p_{c}$, we find that the bands corresponding to correctable errors remain well-defined and non-overlapping; our perturbation theory remains well-defined up to at least second order below the critical window $ w_{c} - w  \sim \sqrt{N}$. 

As the weight of the typical errors~\cite{typicalnote} 
increases, the spectral weight in $\rho^{\l  p \r}$ is dominated by bands of weight $pN$, as long as $p<p_c$. The number of states in such bands grows with $p$ until at $p=p_c$ it reaches $\sim q^N$ (i.e., the physical Hilbert space dimension) and thus cannot continue to grow. Past this point, typical errors are uncorrectable, and the von Neumann entropy density of $\rho^{\l  p \r}$ reaches that of the maximally mixed state.  

The structure of well-defined bands gives a physical mechanism for why the error correction transition in random codes should saturate the hashing bound, consistent with our numerics, and also explains the finite-size scaling behavior at this transition. We confirm that the hashing bound is saturated by an explicit calculation using the Weingarten calculus (see Sec.~\ref{app:weingarten} of the Supplemental Material \cite{suppmat}). The hashing bound was originally derived for random stabilizer codes \cite{Bennett1996hashing,DiVincenzo1997}; we show here that Haar-random codes also saturate it. 

This band structure also sheds light on the physics for $p > p_c$. The error correction threshold is a transition in the structure of the bands that dominate the probability in $\rho^{\l  p \r}$.  However, the low-$w$ bands remain well isolated from each other across $p_c$ and their character does not change across this transition. In particular, the density matrix projected onto the  bands with weight below some $w^*(p)$ retains logical information even for $p > p_c$, so errors remain correctable after post-selecting on these bands, up to a higher detection threshold $p_d$ (although this post-selection is exponentially unlikely to succeed).  Moreover, for $p_c<p<p_d$ individual eigenvalues in these bands are parametrically larger than the typical eigenvalue of $q^{-N}$, so these bands dominate the R\'enyi entropies $S_\alpha \equiv (1-\alpha)^{-1} \glog_q \mathrm{Tr}(\rho^\alpha)$ with $\alpha >\alpha^*(p)>1$. As $p$ increases, each low-weight band hits its own threshold, until at the detection threshold $p_d$, postselected error correction fails even for $w=0$.  We estimate $p_d$ through an elementary combinatorial argument, and confirm our estimate using an exact duality called a quantum MacWilliams identity \cite{Shor_96,Rains_98,rains99,Cao, Cao_24} (defined originally in classical error correcting codes \cite{MacWilliams_book}). (\autoref{app:mbw} -~\autoref{app:rmt} of the Supplemental Material~\cite{suppmat} contains an extensive discussion of quantum MacWilliams identities that may be of independent interest.)

The rest of this paper is structured as follows. In~\autoref{sect:setup} we specify the codes and information-theoretic diagnostics that will be considered. In~\autoref{sect:micro} we introduce a ``fixed-weight'' noise model that helps us isolate and analyze the physics of an individual spectral ``band.'' We then leverage these results to analyze the standard depolarizing noise model, which can be written as a convex sum over fixed-weight noise models, and compute its threshold, in~\autoref{sect:canonical}. In~\autoref{sect:postselect} we explore the behavior of the spectrum for noise rates exceeding threshold, and compute thresholds for post-selected error correction. Finally, in~\autoref{sect:discuss}, we summarize our results and discuss open questions that they raise. 

\section{Setup}\label{sect:setup}

We will discuss the encoding procedure using a standard technique for studying information-theoretic thresholds for quantum codes. Our objective is to encode $k$ logical qudits in a physical system $Q$ comprising $N$ qudits. To this end, we take a subsystem $L$ containing $k$ qudits, and place each of the $k$ qudits in a generalized Bell pair with a qudit of the ``reference'' system $R$, producing the state $\ket{\Phi_{RL}}$. The other $N-k$ qudits in $Q$ (subsystem $A$) are initialized in an arbitrary pure state $|0\rangle_{A}$. To this initial state, we apply a random encoding unitary $U$ on $Q$ sampled from the Haar measure. The resulting encoded pure state on $RQ$ is 
\begin{equation}\label{eq:encoding-RQ}
    \ket{\Psi_{RQ}} = \left(\mathbbm{1}_R\otimes U_Q\right) \left(\ket{\Phi_{RL} } \ket{0_{A}}\right),
\end{equation}
as shown in~\autoref{fig:noisy-encoding}. Its reduced density matrix on $Q$, denoted $\rho_Q$, is the maximally mixed state on the code space, which is a random $q^k$-dimensional subspace of the physical Hilbert space. 

The density matrix $\rho_{RQ}$ is then sent through an error channel acting on $Q$ to create a corrupted density matrix $\tilde{\rho}_{RQ}$; tracing out $R$ gives $\tilde{\rho}_Q$. The optimal capacity of a (hypothetical) decoder to recover the logical information from $Q$ is quantified by the coherent information~\cite{haarnote}:
\begin{eqnarray}
     \Ic && = \svn{}\l \tilde{\rho}_Q \r - \svn{}\l \tilde{\rho}_{RQ} \r~, \nonumber \\
    {\rm where } \ \    &&     \svn{}\l \rho \r = -\tr{ \rho \glog_q(\rho )} , 
\end{eqnarray}
which ranges from $k$ to $-k$. Maximal quantum coherent information between the physical system, $Q$, and the reference, $R$, indicates the information is safely recoverable~\cite{Schumacher1996,Lloyd1997}. As $Q$ is decohered, $Q$ and $R$ become entangled with the (implicit) environment, reducing \Ic. Once $R$ is maximally entangled with the (implicit) environment, $\Ic = -k$, and none of the originally encoded information is recoverable.

The noise model we are primarily interested in is depolarizing noise that acts independently on each qudit. The single-qudit channel can be represented in Kraus form as
\begin{equation}
    \mc{N}_{i ; p}\l \rho \r = \l 1 - p \r \rho + \frac{p}{q^{2}-1}\sum_{\mu =1 }^{q^{2}-1} E^{\mu}_{i}\rho \, E^{\mu\dagger}_{i}~,
    \label{eqn:dep_channel3-p}
\end{equation}
where $\lcb E^{\mu}_{i} \rcb $ are the $(q^{2}-1)$ nonidentity generalized Pauli operators on qudit $i$, and $E^{0}=\mathbbm{1}$. The global depolarizing channel can be written as $\depol = \bigotimes_{i \in Q} \mc{N}_{i; p }$. If one expands out the product, one arrives at a Kraus decomposition of $\mathcal{N}_{p}$ in terms of global errors $E_{\vec{\mu}} = \otimes_{i \in Q}E_{i}^{\mu_{i}}$. Here the vector $\vec{\mu}$ lists each qudit on which the error acts nontrivially, and the specific operator $\mu_{i}$ acting on that qudit. 

The number of qudits $w(\vec{\mu})$ on which $E_{\vec{\mu}}$ acts nontrivially is called the weight of the error. The global depolarizing channel contains errors of all weights, but it will be helpful in what follows to decompose it in terms of channels $\mathcal{N}^{(w)}$ consisting of errors of fixed weight $w$. There are $\Omega(w) \equiv (q^2 -1)^{w}\binom{N}{w}$ distinct errors of weight $w$, and each of these is equally likely, so we can write 
\begin{equation}
\mathcal{N}^{(w)}(\rho) = \frac{1}{\Omega(w)} \sum_{\{ \vec{\mu} | w(\vec{\mu}) = w\}} E_{\vec{\mu}} \rho E^\dagger_{\vec{\mu}}~.
\end{equation}
Defining $P_w \equiv p^{w} \l 1- p\r^{N-w} \binom{N}{w}$ as the probability of a weight-$w$ error, one can decompose the depolarizing channel as a convex sum of fixed-weight error channels as follows:
\begin{equation}\label{convex}
\depol\l\rho\r = \sum_{w = 0}^N P_w \mathcal{N}^{(w)}(\rho)~.
\end{equation}

\begin{figure}[tb]
\includegraphics[width=\linewidth]{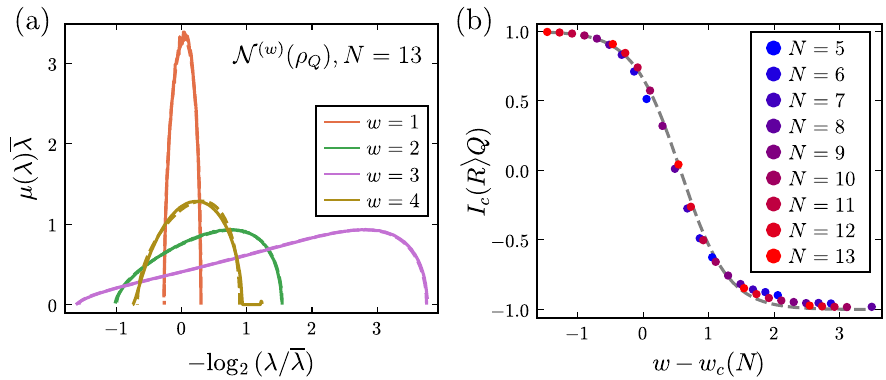}
\caption{Fixed-weight error ensemble, $k=1$ qubit Haar-code. (a) Density of states of the density matrix $\micro{w}(\rho_Q)$, for $N=13, w=1,2$ (below hashing) and $w=3,4$ (above hashing), rescaled to the mean nonzero eigenvalue, $\overline{\lambda} = 1/\min(2^N,2\Omega(w))$ in each band. Solid curves are data aggregated over 467 independent states, while dashed curves are the Marchenko-Pastur distribution with $c=2^{N-1}/\Omega(w)$~\cite{suppmat}. (b) Transition in the coherent information of the fixed-weight error ensemble, with $w_c(N) = p_c N$. Each data point is an average over at least 200 independent states, and error bars are smaller than the markers. The dashed curve, shown to guide the eye, is a fit to a sigmoid function using data points with $N\geq 8$.\label{fig:micro}}
\end{figure}

\section{Fixed-weight error ensemble}\label{sect:micro}

We first discuss the error correction threshold for the family of fixed-weight error channels \micro{w}, as a function of the error weight $w$. As we will see, this ensemble allows for a simpler interpretation of the error correction threshold than the standard local error channel. Moreover, since the error-weight distribution for the local error channel $\depol{}$ is tightly concentrated about its mean (with width scaling as $\sqrt{N}$), one might expect $\micro{pN}$ to be a reasonable proxy for the product of local channels $\depol{}$, at least for some physical quantities.

The channel \micro{w} maps any initial pure state to a mixture of $\Omega(w)$ random pure states. 
Random vectors in a high-dimensional space are approximately orthogonal, so long as the number of random vectors is much smaller than the dimension of the space. 
On the other hand, when the number of random vectors is much larger than the dimension of the space, the resulting mixed state is close to maximally mixed. 

As a first approximation, therefore, \micro{w} maps $\rho_Q$ to a mixed state consisting of $q^k \Omega(w)$ orthogonal vectors if $q^k \Omega(w) \leq q^N$, and to the maximally mixed state otherwise. Meanwhile, $\rho_{RQ}$ is mapped to a state consisting of $\Omega(w)$ orthogonal vectors if $\Omega(w) \leq q^{N+k}$, and to the maximally mixed state otherwise. 
In this approximation, there are three regimes of behavior. First, when $q^k \Omega(w) \leq q^N$, $\tilde{\rho}_Q$ is proportional to a projector onto a $q^k \Omega(w)$-dimensional subspace, and $\tilde{\rho}_{RQ}$ to a projector onto an $\Omega(w)$-dimensional subspace. From the definition of the coherent information, it follows that $I_c(R\rangle Q) = k$. 
Second, when $\Omega(w) \geq q^{N+k}$, both $\tilde{\rho}_Q$ and $\tilde{\rho}_{RQ}$ are maximally mixed, and one has $I_c(R\rangle Q) = - k$ qudits. Finally, for $q^{N-k} \leq \Omega(w) \leq q^{N+k}$, $\tilde{\rho}_Q$ is maximally mixed, but $\tilde{\rho}_{RQ}$ has entropy $S_{\rm vN}(\tilde{\rho}_{RQ}) = \mathrm{log}_q \Omega(w)$.

Expressing these results in terms of the Shannon entropy of the uniform error distribution of the fixed weight error ensemble, we obtain the leading-order ansatz:
\begin{equation}
\label{eq:Ic_fixedweight_ansatz}
    \Ic = \begin{cases} k & H(p) \leq 1 - k/N~, \\
    -k & H(p) \geq 1 + k/N~, \\
    N(1 - H(p)) & \mathrm{otherwise}~,\end{cases}
\end{equation}
where $p=w/N$ and we have defined the entropy of the error distribution 
\begin{equation}\label{eq:H}
    H(p) = -(1-p) \glog_q (1-p) - p~ \glog_q \left[\frac{p}{q^2-1}\right]~,
\end{equation}
and used the relation
\begin{equation}
\Omega(pN) \sim q^{N H(p)}.
\end{equation}
The condition $H(p) \leq 1-k/N$ for perfect information recoverability is the familiar \textit{hashing bound}, which traditionally arises in the context of non-degenerate stabilizer codes~\cite{nondegenerate}.
In the Supplemental Material, we provide an \emph{exact} derivation of Eq.~\eqref{eq:Ic_fixedweight_ansatz} as the leading-$N$ behavior of the coherent information under the full channel $\mathcal{N}_p$, using the Weingarten calculus. For finite-rate codes ($k/N=r=O(1)$), the coherent information decreases over an extensive interval of weights, crossing zero at $p^*$ such that $H(p^*)=1$, while for zero-rate codes ($k=O(1)$), $p^*=p_c \equiv w_c/N \approx 0.189$ (for qubits) marks a transition directly from the ``coding'' to the ``noncoding'' phase.

Above, we oversimplified by treating the channel \micro{w} as mapping each random vector onto a set of strictly orthogonal vectors. In fact, for the basis of orthogonal errors $\{E_{\vec{\mu}}\}$ introduced above, and a basis of $q^k$ orthogonal codewords $\{\ket{\psi^{(n)}}\}$ spanning the support of $\rho_Q$, the pairwise overlap 
$\lv\braket{E_{\vec{\mu}}\psi^{(n)}}{ E_{\vec{\mu}'}\psi^{(n')} }\rv \sim q^{-N/2}$ 
for $\vec{\mu} \neq \vec{\mu'}$ or $n \neq n'$.
Mixed states consisting of ensembles of $\phi$ random vectors in a $d$-dimensional Hilbert space have been considered in the literature~\cite{MarPast,Livan_2018}; the eigenvalue density of these mixed states follows the (appropriately rescaled) Marchenko-Pastur distribution with parameter $c = d/\phi$. We note that the identification of Marchenko-Pastur statistics in the case of Haar random codes remains unproven, and neglects certain matrix element correlations, which are discussed briefly in \cite{suppmat}.

After rescaling to the mean nonzero eigenvalue $1/\min(d,\phi)$, the distributions for parameters $(c, c^{-1})$ with $c < 1$ coincide except for a large number of zero eigenvalues in the latter case (which will not concern us). In particular, in our setting, the ratio of the width to the mean is $\sim \min(\sqrt{c}, 1/\sqrt{c})$, where $c \approx q^N/(q^k \Omega(w))$. As $N \to \infty$, therefore, $c$ remains nonzero and finite only if $w$ is within $O(1)$ of the hashing bound. Outside this transition region, the eigenvalue distribution collapses to a delta function, and the effects of nonorthogonality can be neglected for $(w_c-w)\gg 1$. 

Numerical evidence for these predictions from Haar-random qubit codes, with $k=1$ and $N\leq 13$, is shown in~\autoref{fig:micro}: (a) the eigenvalue density follows the Marchenko-Pastur distribution with $c$ specified above, and (b)~the width of the transition region remains $O(1)$ in $w$ even as $N \to \infty$.

\begin{figure}[bt]
\includegraphics[width=\linewidth]{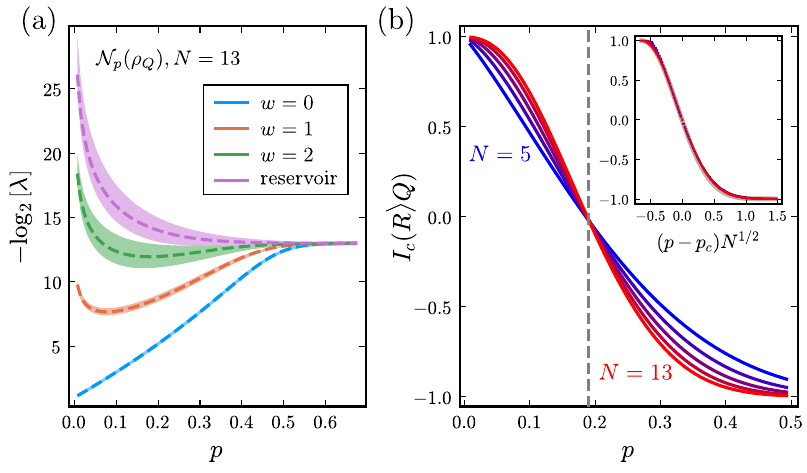}
\caption{Depolarizing error channel, $k=1$ qubit Haar-code. (a) Eigenvalues of the density matrix $\mathcal{N}_p(\rho_Q)$, as a function of depolarization rate $p$. Filled curves are data aggregated from $\approx 100$ samples at $N=13,k=1$. Dashed curves are the ansatz. (b) Coherent information for $k=1$, odd $N$ ranging from 5 to 13. Inset: scaling collapse with $p_c = p_{hash} \approx 0.1893$ for all $N$ in the interval $[7,13]$. Gray curve shows the ansatz for $N=13$.\label{fig:canonical}}
\end{figure}

\section{depolarizing error channel}\label{sect:canonical}

We now return to the depolarizing error channel at error rate $p$, $\depol{}$, which is a convex sum of the fixed-weight channels considered in the previous section [Eq.~\eqref{convex}]. For ease of discussion, we will focus on zero-rate codes, which undergo a single transition at $p=p_c$, but the arguments can be straightforwardly generalized to finite-rate codes.

The eigenvalue structure of $\depol{}(\rho_Q)$ is more complex than that of \micro{w}, but at sufficiently small $p$ they are simply related. Each \micro{w} contributing to $\depol{}$ maps $\rho_{Q}$ onto a mixed state consisting of $q^k\Omega(w)$ random pure states; each of these random states has weight $P_w / (q^k \Omega(w))$ in the ensemble. If we treat all of these states as pairwise orthogonal, we arrive at the following simple description of the eigenvalue spectrum of $\depol{(\rho_Q)}$: it consists of many ``bands,'' each associated with errors of some weight $w$, and consisting of $q^k \Omega(w)$ roughly degenerate eigenvalues of magnitude $P_w/(q^k \Omega(w))$.

As $p \to 0$ this picture becomes asymptotically accurate for the bands with $w < w_c$: $P_w \sim p^w$, so bands corresponding to different $w$ have widely separated eigenvalues and can be treated independently. Even in this regime, errors with larger $w$ do contribute to the spectrum: they fill in the ``leftover'' dimensions of the state space that have not been used by the small-$w$ eigenvalues corresponding to correctable errors, and at low $p$ are well approximated as one more partially filled band with $w=w_c$. We refer to this last band as the ``reservoir''; beyond the error threshold it strongly mixes with neighboring bands, degrading the coherent information. 

To improve upon this approximation and to capture the behavior at larger $p$, we need to account for the fact that the states produced by different Pauli errors are not strictly orthogonal. We treat the effects of nonorthogonality perturbatively~\cite{suppmat}, and find good agreement with the numerically observed eigenvalue density, shown for $N=13, k=1, q=2$ in \autoref{fig:canonical}(a). The effects of this nonorthogonality can be classified into three types: (i)~overlaps between states produced by different errors of the same weight, which we accounted for in the previous section using the Marchenko-Pastur distribution; (ii)~interband overlaps that rigidly shift each band relative to the others, and which can be computed in first-order perturbation theory; and (iii)~interband overlaps that reshape individual bands.  We argue in~\autoref{app:pert_ansatz} of the Supplemental Material that effects of type (iii) are asymptotically subleading in the $N\to \infty $ limit except for bands very close to the critical hashing weight. We have verified that this simple perturbative ansatz consisting of rigidly shifted bands with Marchenko-Pastur-distributed eigenvalues is in good agreement with our numerical results, and show that the Marchenko-Pastur distribution and rigid shifts capture the leading-order physics (\textit{i.e.}, up to parametrically small corrections) for $w_{c}-w\gtrsim \sqrt{N}$ up to second order in \cite{suppmat}.

This picture of well-defined bands, each associated with errors of some weight $w$, implies that the threshold transition at $p_c$ should have width $\sim 1/\sqrt{N}$ for zero-rate codes. This follows because the probability mass of the density matrix $\depol{(\rho_Q)}$ is dominated by errors of weight $w \in (pN - O(\sqrt{N}), pN + O(\sqrt{N}))$. When $|p - p_c| < O(1/\sqrt{N})$, the dominant errors have an appreciable probability of being either correctable or uncorrectable. Therefore, error correction fails with an $O(1)$ probability throughout this regime, and the local error model has a parametrically wider transition region than the fixed-weight error model (though both regions shrink to zero as a function of $p$ when $N \to \infty$). This prediction for the width of the transition region is consistent with numerics, as seen from the scaling collapse of the coherent information as a function of $(p-p_c)N^{1/\nu}$ with $\nu=2$ in \autoref{fig:canonical}(b). This scaling also satisfies Chayes \textit{et al.}'s general bound on finite-size scaling exponents in the presence of disorder~\cite{Chayes1986}, which in our setting reads $\nu\geq 2$. Since the scaling exponent in the fixed-weight error channel is convincingly less than 2 (indeed, the assumption of Marchenko-Pastur statistics suggests $\nu=1$ in that case), we expect the depolarizing channel to saturate this inequality: the transition is broadened solely by the variance of error weights.

\begin{figure}[bt]
\includegraphics[width=\linewidth]{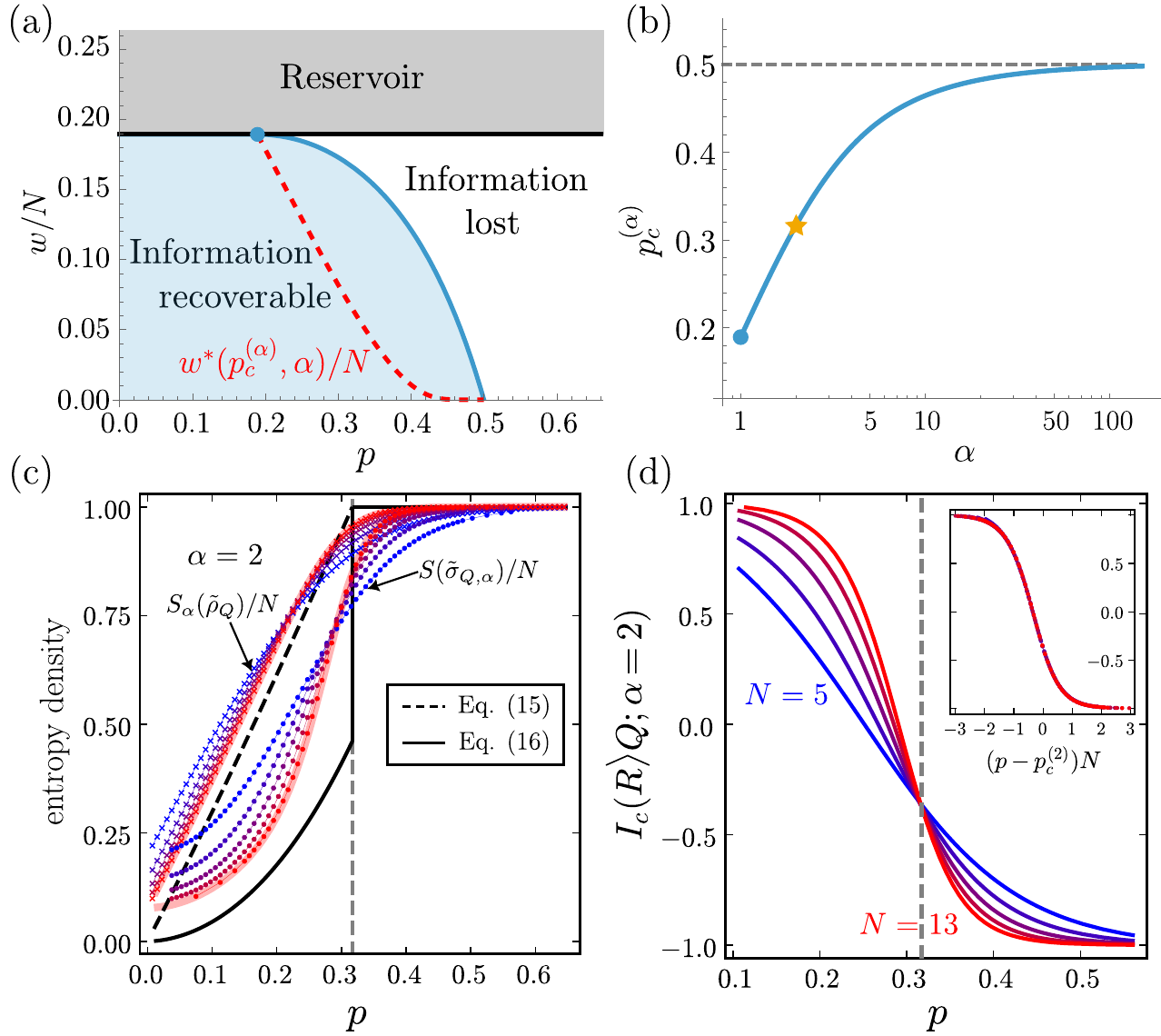}
\caption{(a) Theoretical phase diagram for postselection in a zero-rate qubit Haar-code. Information is recoverable at error rate $p$ by postselecting on band $w$ within the blue-shaded region. Dashed curve shows the threshold under the soft postselection defined in~\autoref{eq:M-alpha}, corresponding to the R\'enyi threshold $p_c^{(\alpha)}$ plotted as a function of $\alpha$ in (b). The star in (b) marks $\alpha=2, p_c^{(2)} = (3-\sqrt{3})/4$. This threshold manifests as both (c) a nonanalyticity in $S_2(\tilde{\rho}_Q)/N$ (x's) and $S_{\rm vN}(\tilde{\sigma}_{Q,2})/N$ (circles) and (d) a transition in the coherent information of $\tilde{\sigma}_{Q,2}$. The main panels show odd system sizes ranging from $N=5$ to $N=13$. In (c), black lines indicate the leading-$N$ behavior for the two quantities (\autoref{eq:renyi} and \autoref{eq:reweight}), while the barely visible red curves are the theoretical calculations for $N=13$. The inset of (d) shows a scaling collapse of the postselected coherent information, with system sizes ranging from $N=7$ to $N=13$.\label{fig:postselect-renyi}}
\end{figure}

\section{Postselection past threshold}\label{sect:postselect}

Above the hashing bound, error correction fails because there are too many typical errors to distinguish: the spectrum of $\rho_Q$ is dominated by states of high $w$ that cannot be associated with specific errors. As shown in~\autoref{fig:canonical}(a), the bands with $w$ near $w_c$ also merge as one increases the error rate past $p_c$. Nevertheless, the bands associated with low-weight errors remain well-defined, and their states do not strongly mix with states produced by uncorrectable errors. This motivates us to consider two forms of postselection, by which quantum information can be recovered past the hashing bound.

\subsection{Hard postselection}
We first consider \textit{hard postselection}: If we define a projector $\Pi_w$ onto the space of states spanned by all codewords acted upon by errors of weight $w$, then the density matrix $\sigma_{p,w} \propto \Pi_w \tilde{\rho}_Q \Pi_w$ projected onto this space may contain only correctable errors \cite{footnote-proj}, even for $p>p_c$. In this regime, postselecting on errors of weight $w$ is exponentially unlikely to succeed, but if it does, then the quantum information remains protected. This projector $\Pi_w$ is not possible to construct in a scalable fashion, but consideration of this postselection is conceptually informative. Note also that when the bands are spectrally separated, this projector is approximately the projection onto one band.

An instructive limiting case is postselecting on the space spanned by errors of weight zero, \textit{i.e.}, on the logical subspace \cite{Turkeshi2024,English2025}, {a common tactic in both near-term experiments and proposed fault-tolerant protocols~\cite{gidney2024}.} An initial logical state can map into this subspace by one of two mechanisms: (i)~the state was not subject to any errors, or (ii)~an arbitrary error occurred, but mapped the initial state to one that was not strictly orthogonal to the logical subspace.  In our model, case~(i) occurs with probability $(1-p)^N$, and case~(ii) with probability $(1-(1-p)^N)q^{k-N}$ (where the factor $q^{k-N}$ is the trace of a random state on the logical subspace, \textit{i.e.}, the probability that a random state will project onto a subspace consisting of $q^k$ states). When $-\glog_q (1-p) \leq 1-k/N$, conditional on postselection succeeding, it is overwhelmingly likely in the limit of large $N$ that no error acted, and the quantum information remains protected. In this sense, the success or failure of postselection provides sharp information on whether an error occurred, and errors are therefore ``detectable''. 

When this inequality fails, errors are not detectable: even if one successfully postselects on the ``no-error'' syndrome, it yields no information about which error acted. 
For $O(1)$  logical qubits encoded in $N$ physical qubits, this detection threshold occurs at $p_d = 1/2$. The argument above is consistent with a more explicit calculation of the detection threshold using a high-low temperature duality, laid out in the Supplemental Material~\cite{suppmat}. 

This argument for the detection threshold readily generalizes to error thresholds upon postselecting for errors of weight $w <w_c$. The probability that an error of precisely weight $w$ acted is $P_w$, while the fraction of states that are in the band of weight $w$ scales as $f_w \equiv q^{k-N} \Omega(w)$. These curves cross when 
\begin{equation}
-\frac{w}{N}  \mathrm{log}_q \frac{p}{q^2-1} - \left(1-\frac{w}{N}\right) \mathrm{log}_q (1-p) = 1-k/N,
\end{equation}
which can be solved for $p$ to determine a postselected error threshold. This postselected threshold is plotted for zero-rate qubit codes in~\autoref{fig:postselect-renyi}(a) (blue curve); it interpolates smoothly between the hashing weight $w_c$ (for $p \leq p_c$) and the detection threshold at $p_d = 1/2$. Note that this postselected threshold is accompanied by a nonanalytic change in the probability that postselection succeeds: below (above) the postselected threshold this probability scales as $P_w$ $(f_w)$. 

\subsection{Soft postselection: R\'enyi reweights}
The perspective of postselected error correction also sheds light on the evolution of R\'enyi entropies with $p$, as we now discuss. Recall that the $\alpha-$Renyi entropy is defined as $S_\alpha(\tilde\rho_Q) \equiv (1-\alpha)^{-1} \glog_q\left[\mathrm{Tr}(\tilde\rho_Q^\alpha)\right]$. Singularities in this quantity have been widely explored in the literature~\cite{FanBao_24,Su_24,Lyons_24}, {{owing to the tractability of integer $\alpha>1$ in statistical mechanics models of decohered stabilizer codes}}, but a clear operational meaning for these singularities (apart from the von Neumann limit $\alpha\rightarrow 1$) has been elusive. {{In this subsection, we show how a ``soft'' postselection protocol tuned by $\alpha$ offers two information-theoretic interpretations of these singularities.}} 

In place of the ``hard'' projectors onto individual $w$ considered in the previous subsection,
we introduce a family of POVMs $\{F_1, F_2\}$ that reweight each eigenvalue of $\rho_Q$ as follows:
\begin{equation}
\{F_1^{(\alpha)} = M_\alpha^\dag M_\alpha, F_2^{(\alpha)} = \mathbbm{1}_Q - F_1^{(\alpha)}\}
\end{equation}
where
\begin{equation}\label{eq:M-alpha}
    M_\alpha = \sum_{n = 1}^{q^k} \left(\frac{\lambda_n}{\lambda_0}\right)^{(\alpha-1)/2} \dyad{n},
\end{equation}
and $\{\ket{n}\}$ is an eigenbasis for $\tilde{\rho}_Q$, with $\lambda_0$ the leading eigenvalue. This is a valid (albeit impractical to implement) POVM, provided $\alpha \geq 1$, and reduces to a projector onto the $w=0$ band when $\alpha=\infty$. Postselecting on obtaining outcome $1$ yields
\begin{equation}
\tilde{\sigma}_{Q,\alpha} \equiv \frac{M_\alpha \tilde{\rho}_Q M_\alpha^\dag}{\tr{M_\alpha \tilde{\rho}_Q M_\alpha^\dag}} = \frac{\tilde{\rho}_{Q}^\alpha}{\tr{\tilde{\rho}_{Q}^\alpha}}.
\end{equation}
For $\alpha<\infty$ and $p<p_d$, the probability that postselection succeeds, $\tr{\tilde{\rho}_Q^\alpha}/\lambda_0^{\alpha-1} = q^{(\alpha-1)(S_\infty(\tilde{\rho}_Q) - S_\alpha(\tilde{\rho}_Q))}$, is a smooth function of the $\alpha$-R\'enyi entropy, which has two regimes of behavior.
At large $p$, $\tilde\rho_Q$ is close to the identity matrix, so postselecting with Eq.~\eqref{eq:M-alpha} uniformly rescales all the eigenvalues, and $S_\alpha(\tilde\rho_Q) \approx N$. However, for smaller $p$, the rescaled density matrix is dominated by large eigenvalues corresponding to distinguishable errors (which are individually likely but collectively unlikely past $p_c$). At the R\'enyi threshold $p_c^{(\alpha)}$, $\tilde\rho_Q^\alpha$ changes from being dominated by correctable errors to uncorrectable errors.

While our quantitative predictions below will focus on Haar-random codes, this mechanism for the non-analyticity in $S_\alpha(\tilde{\rho}_Q)$ applies qualitatively to any sensible code and error model: (some class of) individual correctable errors are likelier to occur, and hence are associated with larger eigenvalues than individual uncorrectable errors. Biasing towards larger eigenvalues by increasing $\alpha$ thus results in a monotonically increasing threshold $p_c^{(\alpha)}$. {{Monotonicity may be violated in certain degenerate codes~\cite{suppmat}, though it has been conjectured to hold in a large family of stabilizer codes~\cite{Su_24}.}}

In Haar-random codes, one can use the zeroth order ansatz for the spectrum of $\tilde{\rho}_Q$, $\lambda(w) = P_w /( q^{k}\Omega(w))$, to estimate the weight $w^*(p,\alpha)$ which dominates the spectrum of $\tilde \rho_Q^\alpha$~\cite{suppmat}:
\begin{equation}\label{eq:renyi-wstar}
\frac{w^*(p,\alpha)}{N} = \frac{(q^2-1) (p/(q^2-1))^\alpha}{(1 - p)^\alpha  + (q^2-1) (p/(q^2-1))^\alpha}.
\end{equation}
As $p$ increases, so does $w^*(p,\alpha)$, until at $p_c^{(\alpha)}$, the contribution to $\mathrm{Tr}(\tilde \rho_Q^\alpha)$ from errors of weight $w^*(p,\alpha)$ becomes subdominant to the contribution $\sim q^{N(1-\alpha)}$ from uncorrectable errors at larger $w$. For $p > p_c^{(\alpha)}$, therefore, $S_\alpha(\tilde\rho_Q) = N$ is pinned to its maximum value. 

The threshold $p_c^{(\alpha)}$ is determined by a R\'enyi generalization of the hashing bound, $H_\alpha(p) = 1 - k/N$, where
\begin{equation}\label{eq:shannon-renyi}
H_\alpha(p) = \frac{1}{1-\alpha} \glog_q \left[(1-p)^\alpha + (q^2-1)\left(\frac{p}{q^2-1}\right)^\alpha\right]
\end{equation}
is the R\'enyi version of the classical Shannon entropy [\autoref{eq:H}]. This threshold is shown in~\autoref{fig:postselect-renyi}(b), again interpolating between the correction threshold $p_c$ at $\alpha = 1$ and the detection threshold $p_d$ as $\alpha\rightarrow\infty$. A saddle-point approximation, in agreement with a calculation using Weingarten calculus, yields a leading-$N$ behavior of
\begin{equation}\label{eq:renyi}
\frac{S_\alpha(\tilde{\rho}_Q)}{N} = \begin{cases}
H_\alpha(p) + k/N & p < p_c^{(\alpha)} \\
1 & p \geq p_c^{(\alpha)}
\end{cases}.
\end{equation}
We note that for $\alpha=2$ and $\alpha=\infty$, the R\'enyi transition is pinned to $p_c^{(\alpha)}$ by the quantum MacWilliams identity, a high-low temperature duality which applies to general codes~\cite{Shor_96}. For these values of $\alpha$, we can also perform an annealed average to obtain the finite-size rounding of $S_\alpha(\tilde{\rho}_Q)$, shown for $\alpha = 2$ in~\autoref{fig:postselect-renyi}(c)~\cite{suppmat}.

So far, we have seen that the R\'enyi entropy $S_\alpha$ changes nonanalytically at $p_c^{(\alpha)}$, and tied this transition operationally to a change in the probability that postselection succeeds. However, the change in the character of $\tilde\rho_Q^\alpha$ also manifests itself as an error-correction threshold for the post-selected state $\tilde\sigma_{Q,\alpha}$. 
For $p > p_c^{(\alpha)}$, $\tilde\sigma_{Q,\alpha}$ is dominated by uncorrectable errors; for $p < p_c^{(\alpha)}$, it is dominated by correctable errors of weight $w^*(p,\alpha) < w_c$. Its (von Neumann) entropy per qubit, in the limit of infinite system size, is therefore
\begin{equation}\label{eq:reweight}
\frac{S_{\rm vN}(\tilde{\sigma}_{Q,\alpha})}{N} = \begin{cases}
H(w^*(p,\alpha)/N) + k/N & p < p_c^{(\alpha)} \\
1 & p > p_c^{(\alpha)}.
\end{cases}
\end{equation}
Unlike the R\'enyi entropy, $S_{\rm vN}(\tilde{\sigma}_{Q,\alpha})$ has a jump discontinuity at $p_c^{(\alpha)}$, which gets rounded out at finite $N$ [\autoref{fig:postselect-renyi}(c)]. 

Intuitively, therefore, one would expect $\tilde\sigma_{Q,\alpha}$ to remain correctable as long as $p < p_c^{(\alpha)}$. This is in good agreement with our numerical findings [\autoref{fig:postselect-renyi}(d)]. We have not been able to derive an analytic ansatz for the coherent information of the post-selected state beyond this saddle-point level: since the postselection protocol involves reweighting the eigenvectors of $\tilde\rho_{RQ}$ by the eigenvalues of $\tilde\rho_Q$, we do not have an explicit expression for the eigenvalues (and therefore the entropies) of $M_\alpha \tilde\rho_{RQ} M_\alpha^\dagger$. However, we observe numerically, and argue heuristically~\cite{suppmat}, that the width of the transition narrows from $\Delta p = O(1/\sqrt{N})$ ($\nu=2$) in the absence of postselection to $\Delta p = O(1/N)$ ($\nu=1$) for strong enough postselection, including $\alpha=2$.

\subsection{Discussion}

In this section we considered two distinct types of postselected transition: directly postselecting on errors of weight $w$, and smoothly reweighting the density matrix by a parameter $\alpha$. In both cases, postselected error correction is possible until some threshold that exceeds the hashing bound $p_c$. Moreover, in both cases, the failure of postselected error correction is accompanied by a nonanalyticity in the probability that postselection succeeds. For the reweighted problem, this nonanalyticity can be interpreted as a singularity in the R\'enyi entropy of the state $\tilde\rho_Q$.

Interestingly, despite these similarities, the two postselection protocols we considered are inequivalent: for a fixed $p$ between the hashing and detection thresholds, one can define $\alpha_{\min}(p)$ and $w_{\max}(p)$ such that postselected error correction succeeds for $\alpha > \alpha_{\min}(p)$ and for $w < w_{\max}(p)$. However, because of the discontinuity discussed in the previous paragraph, $w^*(p,\alpha_{\min}(p)) \neq w_{\max}(p)$: thus, $\alpha$ and $w$ are not simply proxies for one another. Instead, as shown in~\autoref{fig:postselect-renyi}(a), the region in the space of $(p,w/N)$ correctable under $\alpha$-reweighted postselection is strictly smaller than the correctable region under hard postselection; the two thresholds agree only at the hashing bound (no postselection) and detection threshold. This inequivalence is reminiscent of the inequivalence between the microcanonical and canonical ensembles in certain long-range statistical mechanics models~\cite{PhysRevLett.87.030601,Defenu2024,Vicente2025}; it might be interesting to explore whether it also occurs in local quantum codes such as the surface code.

\section{Discussion}\label{sect:discuss}

In this paper we studied how the density matrix $\tilde\rho_Q$ of a random quantum code evolves as it is subjected to depolarizing noise acting independently on all qudits.  We showed that for error rates below the error correction threshold, the spectrum of $\tilde\rho_Q$ consists of well-separated bands.  For all but one of these bands, the eigenstates in each band correspond to correctable errors of some particular Pauli weight. Each such band can be interpreted as the image of the logical space under an error model that only includes errors of a fixed Pauli weight $w$; the full error model is a convex combination of density matrices decohered by such fixed-weight channels. 

In terms of this picture, we were able to derive various spectral and coding properties of these random codes, in particular: (i)~the shape and width of the spectrum of each band, (ii)~the evolution of each band with error rate, and (iii)~the value and finite-size broadening of the error correction threshold. We also showed that the bands corresponding to low-weight errors remain well-defined and correctable between the error correction threshold and the detection threshold: in this regime, we estimated the amount of postselection that is required to make such error correction possible. Finally, we argued that the singularities seen in the R\'enyi entropies past threshold can be related to the success rates for soft postselection into a correctable subspace. This observation gives a concrete information-theoretic meaning to R\'enyi-entropy thresholds that have been extensively discussed in the literature ~\cite{FanBao_24,Su_24,Lyons_24} and further justifies the conjectured monotonicity of these thresholds~\cite{Su_24} (at least for non-degenerate codes).

\subsection{Relation to other works}
A threshold at the hashing bound in the fixed-weight error ensemble is equivalent to the statement that Haar-random codes can perform approximate quantum error correction up to the \textit{quantum Hamming bound}. That is, given a flat error probability distribution over a set of trace orthogonal, unitary errors, Haar-random codes can approximately correct errors when the cardinality of the set is far less than $q^{N-k}$~\cite{ma2025haar}. A rigorous proof of this result appeared nearly simultaneously with our work in Ref.~\cite{ma2025haar}, which introduces the notion of an approximately non-degenerate code with respect to an arbitrary unitary error set. A similar result was proven using a different technique -- which requires only that the code ensemble be a unitary two-design -- in Ref.~\cite{Klesse2007}. We stress that both works are focused on proving a \textit{lower bound} on the fixed-weight/fixed-cardinality threshold, whereas we have argued (using Weingarten calculus, random matrix theory, and numerics) that this bound is tight. While this may seem intuitive on the grounds that \textit{exact} error correction cannot be performed using \textit{exactly} non-degenerate codes past the Hamming bound~\cite{ekert1996}, it may also come as a surprise that, in terms of threshold, Haar-random codes cannot outperform random stabilizer codes.

The connections between our work and Ref.~\cite{ma2025haar} are discussed in detail in the Supplementary Material~\cite{suppmat}. There, we also elucidate the connections between band structure, approximately non-degenerate quantum codes, and the Knill-Laflamme conditions. In brief, when a relative bandwidth remains parametrically small, the conditions for forming an approximately non-degenerate quantum code are satisfied.

\subsection{Generalizations}
In this work, we have presented an analytically tractable minimal model of a decodability/mixed-state phase transition. While the analysis was facilitated by the simplicity of the code ensemble (Haar-random, with no notion of locality) and the noise model (uncorrelated, depolarizing noise), the insights gleaned from our work have both qualitative and quantitative implications for more structured models.

One avenue for extending this work is by modifying the noise model. One analytically tractable noise model is fully heralded, or erasure, errors. The spectrum of a state with $m$ erased qubits admits a straightforward interpretation: to first approximation, like the fixed-weight error ensemble, it obeys a Marchenko-Pastur distribution~\cite{Page1993,Zyczkowski2001,Znidaric2007}, but now with $d\rightarrow q^{N-m}, \phi \rightarrow q^{m + k}$. The threshold and finite-size rounding of the transition, derived using Weingarten calculus in~\autoref{app:erasures} of the Supplemental Material~\cite{suppmat}, are consistent with prior numerical work~\cite{Gullans2021}.

A more challenging avenue is the introduction of correlated noise. This would break a central assumption of our perturbation theory, which states that weights are drawn from Bernoulli trials. Organization into bands around which a perturbation theory could be built is still plausible if certain patterns of errors are dominant, but these bands cannot be organized as simply by many-body weight.

Yet another direction for study is to reintroduce structure into the code ensemble, to determine which aspects of the picture we have presented can generalize to other codes. A simple example are eigenstate codes, where Haar random states are supplanted by energetically equivalent eigenstates of a chaotic Hamiltonian \cite{Brandao2019}. Here the Eigenstate Thermalization Hypothesis (ETH) \cite{Srednicki_ETH,Deutsch_ETH} suggests that our perturbation theory methods could be applied directly (though the implications for thresholds are not entirely clear). 

More nontrivial are local stabilizer codes, such as the toric code, or non-stabilizer codes such as nonabelian topologically ordered states~\cite{Chen2025a,lo2025universalquantumcomputations3,Jing2025,Sala2025}. In the toric code specifically, there is also evidence from statistical mechanics mappings for a series of $\alpha$-dependent singularities in the R\'enyi entropy $S_\alpha(\tilde\rho_Q)$, with the thresholds increasing with $\alpha$ as in random codes. On the other hand, in the toric code and other topological codes, not all errors of a given weight $w$ are equivalent, and the eigenvalue distribution in the large $N$ limit will acquire structure not captured by random matrix theory. Nevertheless, the broad picture of the spectrum splitting into bands may still apply, with the proviso that (for stabilizer codes with degeneracy), one should think of bands as organized by \textit{syndromes}, rather than \textit{errors}. In the Supplemental Material, we develop this perspective by considering the band structure of well-known small stabilizer codes~\cite{suppmat}. The utility of this approach is less clear in the large $N$ limit, as both the eigenvalue multiplicity within bands and the band dispersion become nontrivial to calculate. In light of these challenges, understanding whether some notion of universality applies to the eigenvalue spectra of generic non-degenerate decohered codes (and further exploring the broader similarities in decohered codes with degeneracy) is an interesting task for future work.

 \begin{acknowledgements}
The authors thank Yimu Bao, Xiangyu Cao, Sam Garratt, Michael Gullans, Jake Hauser, Aleksander Kubica, Ethan Lake, Anasuya Lyons, Adam Nahum, Abhinav Prem, Tibor Rakovszky and Curt von Keyserlingk for helpful discussions. We also thank Dongjin Lee for bringing Ref.~\cite{Klesse2007} to our attention. J.~A.~J. was supported by the National Science Foundation Graduate Research Fellowship Program under Grant No. DGE-2039656. D.A.H. was supported in part by NSF QLCI grant OMA-2120757. Z.~W. acknowledges support from the US National Science foundation (NSF) Grant Number 2201516 under the Accelnet program of Office of International Science and Engineering (OISE). Any opinions, findings, and conclusions or recommendations expressed in this material are those of the author(s) and do not necessarily reflect the views of the National Science Foundation. This material is based upon work
supported by the U.S. Department of Energy, Office of Science, National Quantum Information Science Research Centers, Co-design Center for Quantum Advantage (C${}^2$QA) under contract number DE-SC0012704 (theoretical work by S.G.).The simulations presented in this article were performed on computational resources managed and supported by Princeton Research Computing, a consortium of groups including the Princeton Institute for Computational Science and Engineering (PICSciE) and the Office of Information Technology's High Performance Computing Center and Visualization Laboratory at Princeton University. We gratefully acknowledge their hard work, maintaining and improving these computational resources.
 \end{acknowledgements}



\bibliography{Biblio/Enumerators,Biblio/Miscellaneous, Biblio/Stat_Mech_Mappings, Biblio/Nishimori_MLD,Biblio/footnotesv2, Biblio/Weingarten,Biblio/Topological, Biblio/MixedPhases}

\pagebreak

\widetext

\newpage

\input{Supmat-Revised-Clean}

\end{document}

%% file: Supmat-Revised-Clean.tex
\newpage
\onecolumngrid
\makeatletter
\begin{center}
\textbf{\large Supplemental Material: 
Spectral properties and coding transitions of Haar-random quantum codes}

\vspace{3mm}
Grace~M.~Sommers,\textsuperscript{1} J.~Alexander~Jacoby,\textsuperscript{1,2} Zack~Weinstein,\textsuperscript{3,4} David~A.~Huse,\textsuperscript{1}
and Sarang~Gopalakrishnan\textsuperscript{5} 

\vspace{2mm}

\textsuperscript{1}\textit{\small Department of Physics, Princeton University, Princeton, New Jersey 08544, USA}

\textsuperscript{2}\textit{\small Max Planck Institute for the Physics of Complex Systems, 01187 Dresden, Germany}

\textsuperscript{3}\textit{\small Department of Physics, University of California, Berkeley, California 94720, USA}

\textsuperscript{4}\textit{\small Department of Physics and Institute for Quantum Information and Matter, California Institute of Technology, Pasadena, California 91125, USA}

\textsuperscript{5}\textit{\small Department of Electrical and Computer Engineering, Princeton University, Princeton, New Jersey 08544, USA}

\makeatother


\makeatother

\end{center}
\setcounter{section}{0}

\setcounter{equation}{0}
\setcounter{figure}{0}
\setcounter{table}{0}
\setcounter{page}{1}
\makeatletter
\counterwithout{equation}{section}
\renewcommand{\theequation}{S\arabic{equation}}
\renewcommand{\thefigure}{S\arabic{figure}}

The Supplemental Material is organized as follows.
\begin{itemize}
    \item \autoref{app:micro} provides additional detail on the spectral statistics of the density matrix corrupted by fixed-weight errors (``microcanonical channel''). We also comment on connections to the Knill-Laflamme conditions and the proof that Haar-random codes approximately satisfy the quantum Hamming bound.
    \item \autoref{app:pert_ansatz} is a self-contained discussion of the perturbation theory ansatz for the local depolarizing (``canonical'') channel, including the mean eigenvalue in each band and the broadening of each band.
    \item We elaborate upon the soft postselection protocol in~\autoref{app:soft}.
    \item \autoref{app:weingarten} provides an alternative derivation of the hashing bound and R\'enyi thresholds using the Weingarten calculus.
    \item \autoref{app:erasures} uses the Weingarten calculus to derive the threshold and critical behavior of Haar-random codes subject to heralded, or erasure, errors.
    \item The next three sections offer a complementary perspective on the canonical channel and its various thresholds using weight enumerator polynomials and the quantum MacWilliams identity.
    \begin{itemize}
        \item \autoref{app:dual} gives a dual definition of the depolarizing channel.
        \item \autoref{app:mbw} is a self-contained introduction to the many-body weight distribution, weight enumerator polynomials, and the quantum MacWilliams identity.
        \item \autoref{app:macw} discusses the consequences of the quantum MacWilliams identity for the detection and R\'enyi-2 thresholds.
        \item \autoref{app:rmt} derives an exact formula for the Haar-averaged weight enumerator polynomials and demonstrates its consistency with the perturbation theory ansatz.
    \end{itemize}
        \item \autoref{app:stabilizer} elaborates on the band structure of stabilizer codes and its connection to maximum likelihood decoding, including small representative examples.
\end{itemize}

\section{Spectral statistics in the microcanonical channel}\label{app:micro}
\subsection{Identification of Microcanonical Eigenvalue Distributions with Marchenko-Pastur Statistics}
The positive semi-definite square matrix $W = A A^\dag$, where $A$ is an $d \times \phi$  matrix with complex independent and identically distributed (i.i.d.) Gaussian entries, is said to belong to the Wishart (anti-Wishart) ensemble for $d<\phi$ ($d>\phi$). In the former case, $W$ is full-rank, while in the latter, $W$ has $d-\phi$ zero eigenvalues. Defining $c = d/\phi$ as in the main text, in the limit $d,\phi \rightarrow\infty$, the distribution of eigenvalues of $W/\tr{W}$ converges to the Marchenko-Pastur distribution~\cite{MarPast,Livan_2018}:
\begin{equation}
\mu_{d,\phi}(\lambda) \rightarrow 
\begin{cases}
\left(1-\frac{1}{c}\right)\delta(0) + \phi \mu_{c}(\phi \lambda) & c \geq 1 \\
\phi \mu_c(\phi \lambda) & c < 1 
\end{cases}, \quad \mathrm{where} \quad  \mu_c(x)= \frac{1}{2\pi x} \sqrt{\left(x_+ - x\right)\left(x - x_-\right)}, \qquad x_\pm = (1\pm c^{-1/2})^2.
\end{equation}

Now let us apply this to the eigenvalues of the microcanonically corrupted density matrix $\micro{w}(\rho_{RQ})$, which is a $q^{N+k} \times q^{N+k}$ positive semi-definite matrix with trace 1 and $\min(q^{N+k},\Omega(w))$ eigenvalues. Assigning an arbitrary numbering of weight-$w$ errors $E_1, E_2,...,E_{\Omega(w)}$, we can define the $q^{N+k} \times \Omega(w)$ matrix $A$  with entries
\begin{equation}
    A_{ij} \equiv \frac{1}{\sqrt{\Omega(w)}} \left(E_j \ket{\Psi_{RQ}}\right)_i
\end{equation}
so that $\micro{w}(\rho_{RQ}) = A A^\dag$. The matrix elements $A_{ij}$ are neither Gaussian not exactly independent. However, the Marchenko-Pastur scaling form holds for any i.i.d. elements with zero mean and finite variance, and while the $A_{ij}$ are not in fact i.i.d., we find in practice that the correlations are negligible enough for the spectrum of $\micro{w}(\rho_{RQ})$ to be well-modeled by $\mu_{d,\phi}$ with $d=q^{N+k}, \phi = \Omega(w)$. An intuitive perspective as to why one might expect the correlations not to matter is that the random unitary completely scrambles errors: $E_{a}U_{Q}\ket{\Phi_{RL}\otimes 0_{A}}=U_{Q}\l U^{\dagger}_{Q}E_{a}U_{Q}\ket{\Phi_{RL}\otimes 0_{A}}\r $. Alternatively, one can think of ``absorbing '' $E_{a}$ into the Haar measure since it is unitary and the (unimodular) unitary group's Haar measure is also right invariant. Similarly to $\micro{w}(\rho_{RQ})$, $\micro{w}(\rho_Q)$ is a $q^{N} \times q^{N}$ positive semi-definite matrix with trace 1 and $\min(q^N, q^k \Omega(w))$ nonzero eigenvalues, and its distribution closely follows $\mu_{d,\phi}$ with $d=q^N, \phi = q^k \Omega(w)$. 

In~\autoref{fig:micro}(a), we rescaled the distribution by the mean nonzero eigenvalue, which for $\micro{w}(\rho_Q)$ is $\overline{\lambda} = 1/\min(2^N, 2\Omega(w))$. That is, we defined the vertical axis as $c \mu_c(\phi \lambda)$ for $c>1$ and $\mu_c(d\lambda) = c \mu_{1/c}(\phi\lambda)$ for $c\leq 1$.\footnote{This normalization was chosen so that the p.d.f. integrates to 1 for both $c>1$ and $c<1$, without including the delta function weight at zero.} Focusing on $c>1$ without loss of generality, the relative width of the distribution is therefore 
\begin{equation}
    x_+ - x_- = 4/\sqrt{c} \sim \begin{cases}
        \sqrt{\frac{\Omega(w)}{q^{N+k}}} & \micro{w}(\rho_{RQ}) \\
        \sqrt{\frac{\Omega(w)}{q^{N-k}}}& \micro{w}(\rho_Q)
        \end{cases}
\end{equation}
tending to zero as $N\rightarrow\infty$ as long as $c > 1$. This justifies our use of the mean eigenvalue in the ansatz for $\Ic$ (\autoref{eq:Ic_fixedweight_ansatz}), outside an interval of width $O(1)$ weight. Motivated by the form of $c$, we define $w_{c; RQ} \in \mathbb{N}$ as the smallest $w$ such that $\sum_{w' = 0 }^{w}\binom{N}{w' }\l q^{2}-1\r^{w'} \geq q^{N+k}$. Similarly, $w_{c; Q}$ is the smallest $w$ such that $\sum_{w' = 0 }^{w}\binom{N}{w' }\l q^{2}-1\r^{w'} \geq q^{N-k}$.

\color{black}
\subsection{Connection to Proof of Hamming Bound and Knill-Laflamme Conditions}\label{app:kl}
It was recently shown in a nearly simultaneous work \cite{ma2025haar} that Haar random codes attain the quantum Hamming bound, approximately. That is, for unitary error sets $\lv\mc{E}\rv \ll q^{N-k}$, where $P(E)$ is constant for all $E\in \mc{E}$, the Haar code remains an \textit{approximate non-degenerate code}: all error corrupted codewords remain approximately orthogonal to one another. This is defined in terms of isometry
\begin{equation}
    \mc{I}= \sum_{i=1}^{q^k}\sum_{a=1}^{N_{E}} E_{a}\ket{\psi_{i}}\bra{i,a},
\end{equation}\label{eq:encoding_isometry}
which maps a particular choice of logical state and error into the physical Hilbert space. When the singular values, $\lcb \sigma_{i} \rcb$ are all $1$, the error corrupted codewords remain exactly orthogonal. This is loosened by demanding deviations remain bounded: $\lcb \sigma_{i}\rcb \subset \lb 1- \delta, 1+\delta  \rb$. This definition is closely related to the microcanonical bandwidth as can be seen by examining $\mc{I}\mc{I}^{\dagger}$:
\begin{equation}
    \mc{I}\mc{I}^{\dagger} = \sum_{a=1}^{N_{E}}\sum_{i=1}^{q^{k}} E_{a}\ket{\psi_{i}}\bra{\psi_{i}}E_{a}^\dag = \lv \mc{E}\rv q^{k}\times \tilde{\rho}_{Q}^{(w)}.
\end{equation}\label{eq:encoding_isometry2}
Demanding the approximate orthogonality is therefore equivalent to demanding eigenvalues of $\tilde{\rho}_{Q} \times \lv\mc{E}\rv q^{k} $ be bounded on the interval $ \lb \l 1- \delta\r^{2}, \l 1+\delta \r^{2} \rb$. Examining the Marchenko-Pastur distribution conjectured in this work, we may identify $\delta = c^{-1/2}\sim \l \frac{\lv \mc{E}\rv}{q^{N-k}}\r^{1/2}$.

This definition can also be related to the Knill-Laflamme conditions \cite{Knill_2000}, which demand that $\braket{E_{a}\psi_{i}}{E_{b}\psi_{j}}=  \delta_{i,j} M_{a,b}$ where $M_{a,b}$ does not depend on $i,j$. The exact orthogonality of the corrupted codewords is clearly a stronger condition, since 
\begin{equation}
    \mc{I}^{\dagger}\mc{I} = \sum_{ia,jb} \ket{ia}\braket{E_{a}\psi_{i}}{E_{b}\psi_{j}}\bra{jb} \xrightarrow[ ]{{\rm exact\  orth.}}\delta_{ij}\delta_{ab}
    \label{eq:kl}
\end{equation}
so that $M_{a,b}$ is not only constant, but a delta function. Adopting the approximate orthogonality conditions, we can show that the Knill-Laflamme conditions are approximately satisfied, with violations upper bounded by $\l 1+\delta\r^{2}-1$. Beginning in the spectral basis of $\mc{I}^{\dagger}\mc{I} $, \autoref{eq:kl} can be rewritten as the identity plus an operator-norm bounded perturbation, $V$ with $\norm{V} \leq \l1+\delta\r^{2}-1$. Since the identity and operator norm are invariant under a change of basis, Eqn.~\ref{eq:kl} must also be the sum of the identity (which manifestly obeys Knill-Laflamme) and an operator-norm bounded, Hermitian perturbation $V'$ (since both sides of the equation are manifestly Hermitian). This operator norm, in turn, upper bounds all off-diagonal matrix elements and therefore upper bounds the violations of the Knill-Laflamme conditions.
\color{black}

\section{Perturbative Calculations}\label{app:pert_ansatz}

In the main text, we developed a picture of the fixed error-weight transition as occurring when the number of errors becomes sufficiently large that each error does not map the codespace to sufficiently  distinguishable subspaces. In the case of the local depolarizing channel, we must account for a non-trivial distribution of error probabilities. Ultimately, the relevant question in the depolarizing error ensemble is when a \textit{typical} error is no longer correctable; as opposed to the fixed weight error ensemble at $w$ where all weight-$w$ errors are typical, in the depolarizing ensemble a central limiting window of errors of width $N^{1/2}$ around mean weight $w = pN$ will be typical. The smoothing of this transition can be accounted for in an approach that resembles perturbation theory in the overlaps of error corrupted states. This pertubation theory is developed by hierarchically orthogonalizing new subspaces introduced by errors of weight $w$ to the subspaces from lower weight errors. These subspaces correspond to bands. Within the diagonal blocks, the average eigenvalue can be calculated (\autoref{app:MSA}) and we can then show the off-diagonal matrix elements provide only small corrections in the relevant parameter regimes (\autoref{app:PCorrections}). We refer to the computation of the average eigenvalue in the blocks as the ``mean-shift ansatz.''

\begin{figure}[h]
    \centering
    \includegraphics[width=0.85\linewidth]{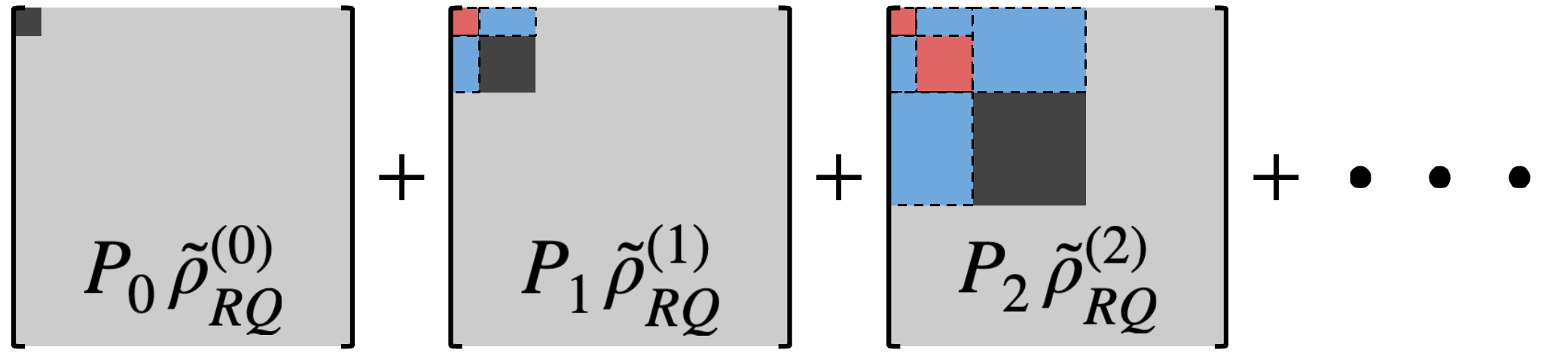}
    \caption{Block structure of the decohered density matrix in the hierarchy orthogonalized block basis with colors corresponding to terms in~\autoref{eqn:pertblock}.}
    \label{fig:perturbation_block_struct}
\end{figure}

We remind the reader of the notation $\tilde{\rho}_{RQ}^{\l w \r} \equiv \micro{w}\l \rho_{RQ}\r $ for the fixed-weight decohered density matrix; and additionally notate the depolarized density matrix $ \tilde{\rho}_{RQ}^{\l p \r} \equiv\sum_{w} P_{w}\tilde{\rho}^{\l w \r}_{RQ}$. We then define 
\begin{equation}
    \mc{H}_{RQ}^{\l w \r} \equiv {\rm support}\l \tilde{\rho}_{RQ}^{\l w\r}\r \setminus \l \cup_{w'=0}^{w-1}\mc{H}^{\l w'\r}_{RQ}\r
\end{equation}
where\footnote{An exactly analogous construction holds with $Q$, where $\mc{H}_{Q}^{\l 0 \r } = {\rm span }\l  \lcb \ket{\psi^{(n)}}\rcb \r$ (the logical space); all band multiplicities are augmented by a factor of $q^{k}$ and we use $w_{c;Q}$ as defined in the previous section.} $\mc{H}_{RQ}^{\l 0 \r } = {\rm span }\l  \ket{\Psi_{RQ}} \r$. Generally, ${\rm dim}\l \mc{H}_{RQ}^{\l w \r}\r = \Omega(w)$ for $w < w_{c}$ and ${\rm dim}\l \mc{H}_{RQ}^{\l w \r}\r = 0$ for $w>w_{c; RQ}$.\footnote{Additionally, we define $\Omega\l w_{c;RQ} \r \equiv{\rm dim}\l \mc{H}^{\l w_{c;RQ}\r}_{RQ}\r \leq \binom{N}{w_{c;RQ}}\l q^{2} -1 \r^{w_{c; RQ}} $, an exception for which there is no closed form for the multiplicity. The partially filled band associated to $\mc{H}^{\l w_{c;RQ}\r}$ has no clear interpretation in terms of errors of a given weight, and we shall refer to it as the ``reservoir.''} To each one of these spaces, labeled by $w$, is associated a band and an orthogonal projector $\pi^{\l w \r}_{RQ}$. We emphasize that projectors $\pi^{\l w\r}_{RQ}$ should be treated like a scalar $0$ when $w>w_{c;RQ}$. Noting that we can resolve the identity as $\mathbbm{1}_{RQ}= \sum_{w=0}^{w_{c;RQ}} \pi^{\l w \r}_{RQ}= \sum_{w=0}^{N} \pi^{\l w \r}_{RQ}$, the density matrix can be written as
\begin{equation}\label{eqn:pertblock}
    \tilde{\rho}^{\l p \r}_{RQ} = \sum_{w, w_{l}, w_{r} = 0 }^{N} P_{w} \, \pi^{\l w_{l}\r}_{RQ}  \tilde{\rho}_{RQ}^{\l w\r} \pi^{\l w_{r}\r}_{RQ}  = \sum_{w} P_{w}\lb \pi^{\l w \r}_{RQ}  \tilde{\rho}^{\l w \r}_{RQ} \pi^{\l w \r}_{RQ} + {\color{red}\sum_{w_{lr} = 0 }^{w-1} \pi^{\l w_{lr}\r}_{RQ} \tilde{\rho}^{\l w \r}_{RQ} \pi^{\l w_{lr} \r}_{RQ}} + {\color{blue}\sum_{w_{l} \neq w_{r}}^{w} \pi^{\l w_{l}\r}_{RQ} \rho^{\l w \r}_{RQ} \pi^{\l w_{r} \r}_{RQ}}\rb
\end{equation}
where terms are organized in block fashion per Fig. \ref{fig:perturbation_block_struct} with corresponding colors.
There are three broad types: (1) The black terms are the microcanonical/fixed-weight density matrices, $\tilde{\rho}^{\l w \r}_{RQ}$, projected into their corresponding sector $\mc{H}^{\l w \r}_{RQ}$. {\color{red} (2)} The red weight-block diagonals are organized into two parts which we will treat separately: the diagonals, which are positive definite and contribute mean shifts to the bands, and the off-diagonals which are only relevant to bands near $w_{c;RQ}$ for $p\gtrsim p_{c}$. {\color{blue}(3)} The blue off block-diagonal terms, similarly to the off-diagonal ${\color{red} (2)}$ terms, will not matter except to bands near $w_{c;RQ}$ for $p\gtrsim p_{c}$.

\subsection{Mean-shift Ansatz}\label{app:MSA}
To understand the mean shifts (diagonals of ${\color{red} (2)}$ and effects of projection of $(1)$) we compute the traces of $P_{w'}\tilde{\rho}_{RQ}^{\l w'\r }$ on different $\mc{H}^{\l w\r}_{RQ}$ where $w' >w$:
\begin{eqnarray}\label{eqn:od_trace}
    P_{w'}\, {\rm Tr}_{\mc{H}^{\l w \r}_{RQ}} \lb \rho_{RQ}^{\l w' \r}  \rb &=& \frac{P_{w'}}{\Omega\l w'\r}\sum_{\lcb \vec{\mu} \vert w(\vec{\mu}) = w' \rcb} {\rm Tr}\lb \pi^{\l w \r}_{RQ}  \ket{E_{\vec{\mu}}\Psi_{RQ}}\bra{E_{\vec{\mu}}\Psi_{RQ}} \rb = \frac{P_{w'}}{\Omega\l w'\r}\sum_{\lcb \vec{\mu} \vert w(\vec{\mu}) = w' \rcb}  \Omega\l w \r \l q^{-\l N + k\r} + ...\r \nonumber \\
    &=& \frac{P_{w'}}{\Omega\l w'\r}\times \Omega\l w'\r \Omega\l w\r q^{-\l N+k\r }  =  \Omega\l w \r   \frac{P_{w'}}{q^{ N+k }},
\end{eqnarray}
where we have estimated the average inner product squared between a basis vector of $\mc{H}^{\l w \r}_{RQ}$ and $E_{\vec{\mu}}\ket{\Psi_{RQ}}$ with $\vec{\mu}$ weight-$w'$ as that of two Haar-random vectors (at first order in $q^{-\l N + k\r}$-- subleading terms are undoubtedly present as a consequence of the orthogonalization). Thus the $\Omega\l w \r$ eigenvalues in band $w<w'$ are shifted on average by $\delta \lambda^{\l  w \r}_{{\rm avg.} ; RQ} =   \sum_{w'>w}P(w') q^{- \l N+k \r}$. Similarly, we can use the trace preservation property of the fixed-weight channel--- \textit{i.e.}, that $\sum_{w'\leq w}{\rm Tr}_{\mc{H}^{\l w' \r}_{RQ}}\lb \rho^{\l w\r }_{RQ}\rb = 1 $ --- to compute the trace of $P_{w}\tilde{\rho}_{RQ}^{\l w\r }$ on $\mc{H}^{\l w \r}_{RQ}$:
\begin{equation}\label{eqn:d_trace}
     {\rm Tr}_{\mc{H}^{\l w\r}}\lb P_{w} \rho_{RQ}^{\l w\r} \rb = P_{w} \l 1 - \sum_{w'<w} {\rm Tr}_{\mc{H}^{\l w' \r}_{RQ}}\lb \rho^{\l w\r }_{RQ}\rb \r = P_{w} \l 1 - \sum_{w'<w} \Omega\l w'\r q^{-\l N + k\r}\r,
\end{equation}
were we have simply substituted the result of results of \autoref{eqn:od_trace} into the trace preservation relation. This yields an average shift of $\delta\lambda^{\l w \r}_{{\rm avg.} ; RQ } =  -P_{w}\sum_{w'<w}\Omega\l w' \r q^{-\l N +  k\r}/\Omega(w)$. 

Putting together the positive and negative shifts we find that the average eigenvalue of $\tilde{\rho}_{RQ}^{p}$ in band $w\leq w_{c;RQ}$ is

\begin{equation}\label{eqn:msa}
    \lambda^{\l w \r}_{{\rm avg.};RQ} = \frac{P_{w}}{\Omega\l w \r}\l 1 - \sum_{w'< w } \frac{\Omega\l w' \r}{q^{N+k}} \r+\sum_{w'>w} \frac{P_{w'}}{q^{N + k }}.
\end{equation}
An example of this perturbative ansatz is shown in the main text in \autoref{fig:canonical}(a), showing strong agreement with numerical evidence. Additionally this ansatz is manifestly trace preserving; this can be seen by applying the identity $\sum_{x=0}^{w_{c}}\sum_{y<x}f(x,y) =\sum_{x=0}^{w_{c}}\sum_{y>x}f(y,x)  $ as 
\begin{equation}
    \sum_{w=0}^{w_{c}} \lambda^{\l w \r}_{{\rm avg.}; RQ} \Omega(w) = 1 +q^{-\l N + k\r}\sum_{w=0}^{w_{c}} \lb \sum_{w'>w}  P_{w'} \Omega\l w \r - \sum_{w'< w}P_{w} \Omega\l w'\r\rb = 1 .
\end{equation}
We note that the above equality relies upon a splitting of terms in the positive-shift sum between those with $w'\leq w_{c}$, which cancel against the negative shifts due to the reindexing identity, and those with $w' >w_{c}$, which contribute to the identity
\begin{equation}
    \sum_{w=0}^{w_{c}} \Omega\l w \r\l \frac{P_{w}}{\Omega\l w \r} +\sum_{w'=w_{c}}^{N}\frac{P_{w'}}{q^{N+k}}\r = \sum_{w=0}^{w_{c}}P_{w} + \sum_{w'>w_{c}}^{N}P_{w'} =1.
\end{equation}
For $\tilde{\rho}_{Q}^{(p)}$, the band multiplicities become $\Omega\l w \r \times q^{k}$ and the critical weight is modified (see definition in \autoref{app:micro}). This modifies the mean eigenvalue to 
\begin{equation}\label{eq:msa-Q}
    \lambda^{\l w \r}_{{\rm avg}; Q} = \frac{P_{w}}{q^{k}\Omega\l w \r}\l 1 - \sum_{w'< w} \frac{\Omega\l w' \r}{q^{N-k}}\r + \sum_{w'>w}\frac{P_{w'}}{q^{N}}.
\end{equation}

\begin{figure}[bt]
\centering
\includegraphics[width=0.8\linewidth]{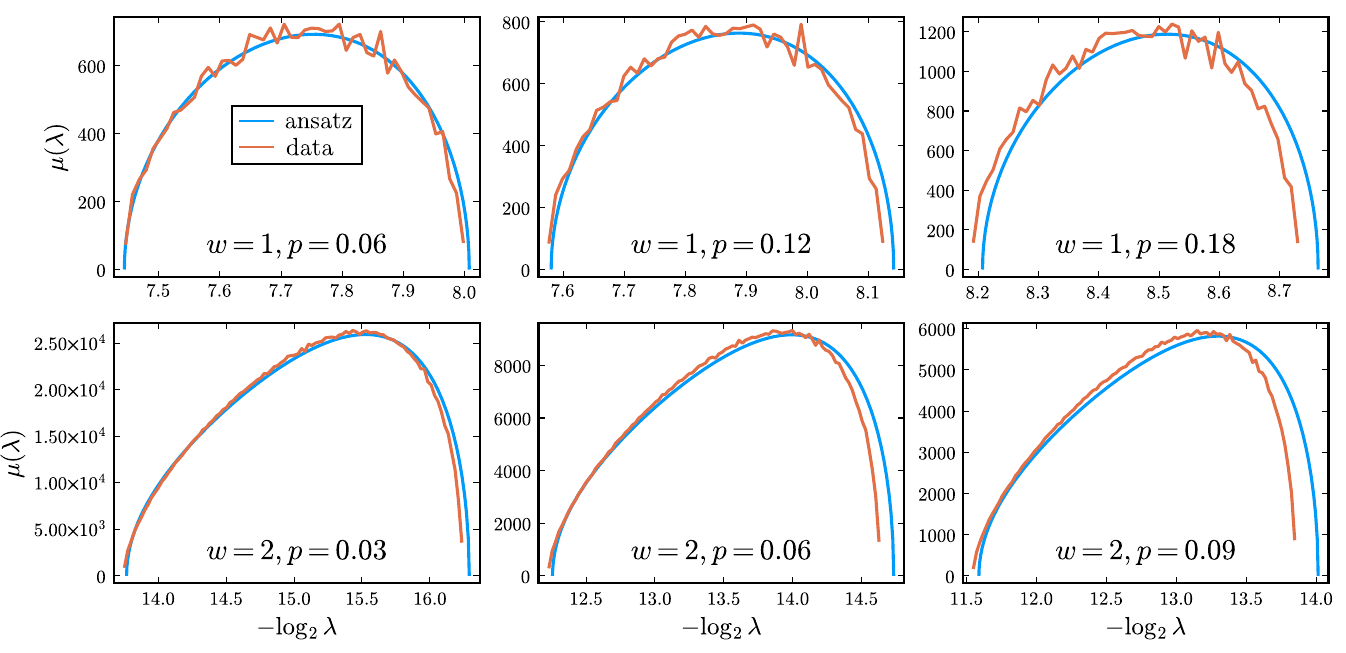}
\caption{The two nontrivial below-hashing bands for $N=13$, $k=1$, $\depol(\rho_Q)$. Top row shows $w=1$ at error rates $p=0.06,0.12,0.18$. Bottom row shows $w=2$ at error rates $p=0.04,0.08,0.12$. Blue curves come from a rigid shift of the microcanonical ansatz, multiplied by $P_w$, while orange curves are the numerical p.d.f. aggregated over 102 samples. \label{fig:canonical-bands}}
\end{figure}

\autoref{fig:canonical}(a) in the main text provides empirical support for this mean shift ansatz, as~\autoref{eq:msa-Q} follows the middle of each band. Further data for the same system (a single logical qubit encoded into $N=13$ physical qubits) are shown in~\autoref{fig:canonical-bands}, demonstrating that at low $p$, adding this mean shift to the Marchenko-Pastur distribution used to model the microcanonical bands (rescaled by a factor of $P_w$) successfully captures the spectrum of the entire canonical band. In the remainder of this section, we consider corrections to this picture and identify the regimes in which they can be neglected.

\subsection{Perturbative Ansatz and Distributions} \label{app:PCorrections}
Now we consider the effects of the interband mixing on the distribution of eigenvalues for $\tilde{\rho}_{RQ}^{\l p \r}$; generalizations to $\tilde{\rho}_{Q}^{\l p\r }$ are straightforward and we will therefore leave them implicit. $P_{w}\tilde{\rho}_{RQ}^{\l w \r}$ has $\Omega\l w \r$ positive-definite eigenvalues supported on  the interval
\begin{equation}
    \mc{I}_{w} = \frac{P_{w}}{\Omega\l w \r} \left[\left(1 - \sqrt{\frac{\Omega\l w \r}{q^{N+k}}}\right)^2, \left(1 + \sqrt{\frac{\Omega \l w \r}{q^{N+k}}}\right)^2\right]
\end{equation}
which has a width of 
\begin{equation}\label{eq:micro-width}
    \Delta\l w \r = \lv \mc{I}_{w}  \rv = \frac{P_{w}}{\Omega\l w \r}\lb\left(1 + \sqrt{\frac{\Omega \l w \r}{q^{N+k}}}\right)^{2} - \left(1 - \sqrt{\frac{\Omega \l w \r}{q^{N+k}}}\right)^{2} \rb =  \frac{4 P_{w}}{\sqrt{\Omega\l w \r q^{N+k}}} = 4 \frac{P_{w}}{\Omega\l w \r} \sqrt{\frac{\Omega\l w \r}{q^{N+k}}}.
\end{equation}
We will argue that this width is robust to the fluctuations due to diagonals of terms ${\color{red} (2) }$ and due to the projection of $\tilde{\rho}^{\l w \r}_{RQ}$ onto $\mc{H}_{RQ}^{\l w \r}$. In the above expression $P_{w}/\Omega\l w \r$ is recognizable as the probability of a given error at weight $w$: 
\begin{equation}\label{eq:p-error}
    \frac{P_{w}}{\Omega(w)} =  \left(\frac{p}{q^2-1}\right)^{w} (1-p)^{N-w} \equiv u^{w}\l 1 - p \r^{N} \qquad {\rm where} \qquad  u \equiv \frac{p}{(q^2-1)(1-p)}.
\end{equation}
We note the monotonicity property of the given error probability: $P_{w'}\Omega\l w\r/P_{w}\Omega\l w' \r = u^{w'-w} <1$ if $w'>w$ for all $u<1 \impliedby p < p_{max} =  1 - q^{-2}$.\footnote{This property can be simply determined from the monotonicity of the function $x/\l 1-x\r$ and the relation $1 = \left. u\rv_{p =   1 - q^{-2}}$.}

\subsubsection{Diagonal Fluctuations}
Now we consider $ \pi^{\l w \r}_{RQ}\l P_{w}\tilde{\rho}^{\l w\r}_{RQ}+ P_{w'}\tilde{\rho}_{RQ}^{\l w'\r}\r \pi^{\l w \r}_{RQ}$ with $w< w'$ and $0<w<w_{c;RQ}$. Per the mean shift ansatz, the diagonal elements of $P_{w'}\pi^{\l w\r}_{RQ} \tilde{\rho}_{RQ}^{\l w' \r}\pi^{\l w \r}_{RQ}$ will exert an average shift on the eigenvalues of $P_{w}\pi^{\l w\r}_{RQ} \tilde{\rho}_{RQ}^{\l w \r}\pi^{\l w\r}_{RQ}$ of $\delta\lambda^{\l w\r}_{{\rm avg.} ; RQ} = P_{w'} q^{-\l N+k\r} $ which can be large compared to $\Delta\l w \r$. However, this is an \textit{average} shift and individual eigenvalue shifts are allowed to fluctuate; they will do so in a central limiting fashion. The RMSD of these (positive) diagonal contributions ${\color{red} (2)}$ is readily calculated and compared to distribution (band)width: 
\begin{equation}
    \sqrt{\sigma^{2}\l \delta \lambda^{\l w \r}_{RQ}\l w' \r \r} = \frac{P_{w'}}{\Omega\l w'\r}\frac{\sqrt{\Omega\l w'\r}}{q^{N+k} } \qquad \quad   \frac{\sigma\l \delta \lambda^{\l w \r}_{RQ}\l w '\r\r}{\Delta\l w \r} \sim \frac{P_{w'} \Omega\l w \r}{P_{w}\Omega\l w'\r}\sqrt{\frac{\Omega\l w'\r}{\Omega\l w \r q^{N+k}}}   = \frac{u^{w'-w}}{\sqrt{\Omega\l w \r}}\sqrt{\frac{\Omega\l w'\r}{q^{N+k}}}.
\end{equation}
Since $u^{w'-w}<1$, $\l\frac{\Omega\l w'\r}{q^{N+k}}\r^{1/2}<1$, and $\l\Omega\l w \r\r^{-1/2}<1$, we see that the fluctuations of the shifts cannot substantially broaden the distribution. These fluctuations can dominate the microcanonical level-spacing structure, however, as can be seen from the ratio with the typical level spacing $\delta\l w \r \sim \frac{P_{w}}{\Omega\l w \r} \l\Omega\l w \r q^{N+k}\r^{-1/2}$:
\begin{equation}
    \frac{\sigma\l \delta \lambda^{\l w \r}_{RQ}\l w '\r\r}{\delta\l w \r} \sim  u^{w'-w} \sqrt{\frac{\Omega\l w \r \Omega\l w' \r}{q^{N+k}}},
\end{equation}
which need not be small. We also must estimate the RMSD of the fluctuations introduced by the projection of $P_{w}\tilde{\rho}_{RQ}^{\l w \r}$ into $\mc{H}^{\l w \r}_{RQ}$ (the negative shifts in the mean shift ansatz):
\begin{equation}
    \frac{\sqrt{\sigma^{2}\l \lambda^{w}_{\rm avg} \r}}{\Delta\l w \r} = \frac{P_{w}}{\Omega\l w \r}  \frac{\sqrt{\sum_{w'<w}\Omega\l w'\r}}{q^{N+k}} \times \frac{\sqrt{\Omega\l w \r q^{N+k}}}{P_{w}} = \l \frac{\sum_{w'<w} \Omega\l w'\r/\Omega\l w \r}{q^{N+k}} \r^{1/2}
\end{equation}
which is manifestly small.

\subsubsection{Off-diagonal Matrix Elements}

Now we define a basis of eigenvectors on the blocks described in the previous section. Namely, projecting the microcanonical density matrices into their corresponding sectors as $P_{w}\pi_{RQ}^{\l w \r}\tilde{\rho}_{RQ}^{\l w \r} \pi_{RQ}^{\l w \r}$ and diagonalizing them, we label the eigenvectors $\lcb \gket{ n^{\l w\r} } \rcb$. The eigenvalue shifts due to the diagonal part of the ${\color{red} (2)}$ terms described in the previous section do not impact the structure of these eigenvectors (since the shifts are simply projections of the density matrix onto these eigenvectors). Now, we compute the typical off-diagonal matrix elements given by terms of type ${\color{red} (2)}$ (\textit{i.e.}, $w = w''$ with $m\neq n$) and ${\color{blue} (3)}$ (\textit{i.e.}, $w \neq w''$), which are of the same magnitude; in both cases, we require $w\leq w'$ and $w''\leq w'$ due to the block structure. We get 
\begin{eqnarray}
    M^{\l w' \r } &\equiv& \lv \bra{n^{\l w \r}} P_{w'}\tilde{\rho}_{RQ}^{\l w' \r}\ket{m^{\l w''\r}}\rv^{2} = \l \frac{P_{w'}}{\Omega\l w'\r}\r^{2} \lv \sum_{\lcb \vec{\mu}\vert w\l \vec{\mu}\r = w' \rcb}  \braket{n^{\l w \r}}{E_{\vec{\mu}}\Psi_{RQ}}\braket{E_{\vec{\mu}}\Psi_{RQ}}{m^{\l w'' \r }}  \rv^{2} \nonumber \\
    &\sim&  \frac{P_{w'}^{2}}{\Omega\l w' \r} \l q^{-2\l N +k \r}   + ...\r \approx \frac{P_{w'}^{2}}{\Omega\l w' \r q^{2\l N + k\r}}
\end{eqnarray}
where we keep only the leading term in the Hilbert space dimension, which is independent of $n^{\l w\r}$ and $m^{\l w''\r}$. Subleading terms appear due to the orthogonalization of the basis vectors to lower-weight blocks. In the above calculation we assume that the sum of the matrix elements due to different errors of weight $w$ is central limiting.

\subsubsection{Second-order pertubation theory of terms of type {\color{blue} (3)}}
We can thus calculate the corrections to $\lambda^{\l w \r}_{n}$ from terms ${\color{blue}(3)}$ coming from $P_{w'}\tilde{\rho}^{\l w' \r}_{RQ}$ which mix $\gket{n^{w}}$ with the band $w''$ in second order perturbation theory:
\begin{eqnarray}
    \delta\lambda^{\l w \r}_{n} &=& {\rm sgn}\l w'' - w\r\sum_{m^{\l w''\r}} \frac{\lv\bra{n^{\l w \r}} P_{w'}\tilde{\rho}_{RQ}^{\l w' \r}\ket{m^{\l w''\r}} \rv^{2}}{ \lv \lambda^{\l w \r}_{n} - \lambda^{\l w''\r}_{m} \rv} \sim \pm \Omega\l w''\r M^{\l w'\r} {\rm min}\l \frac{\Omega\l w\r}{P_{w}}, \frac{\Omega\l w''\r}{P_{w''}}\r \nonumber \\
    &\sim& \pm \frac{\Omega\l w''\r P_{w'}^{2}}{q^{2\l N+k\r} \Omega\l w'\r}{\rm min}\l \frac{\Omega\l w\r}{P_{w}}, \frac{\Omega\l w''\r}{P_{w''}}\r = \frac{\Omega\l w''  \r P_{w'}}{q^{2\l N + k \r}} {\rm min }\l u^{w'-w}, u^{w'-w''}\r.
\end{eqnarray}
This does not change the width of the band since it is a uniform sign. Comparing to the positive first order corrections from $P_{w'} \tilde{\rho}^{\l w' \r}_{RQ}$, we get
\begin{equation}
    \frac{\rm Second \ Order}{ \rm Positive \ First \ Order}=     \frac{\Omega\l w''\r }{q^{ N+k} }{\rm min}\l u^{w'-w}, u^{w'-w''}\r .
\end{equation}
In all cases except $w \sim w' \sim w''\sim w_{c; RQ}$ this ratio remains small. This is because it is manifestly small when $w''$ is not the reservoir and $w'$ must always be larger than $w$ and $w''$, suppressing it exponentially when $w' \sim w'' \sim w_{c; RQ}$ and $w$ is not within $O(1)$ of $w_{c; RQ}$. There is no guarantee that this leading term of the second-order perturbative correction is large compared to the subleading terms of the first order result (which are not straightforwardly computed). So it is of limited practical use, besides evidencing that the pertubation theory remains controlled.

\subsubsection{Second-order pertubation theory of off-diagonal terms of type {\color{red} (2)}}

Now we calculate the typical correction to $\lambda^{\l w\r }_{n}$ from terms of type ${\color{red} (2)}$ due to $P_{w'} \tilde{\rho}^{\l w' \r}_{RQ}$. These calculations appear with both positive and negative signs, but for a typical eigenvalue $\lambda^{\l w \r}_{n}$ the fraction of the spectral weight in band above and below it will not be equal so the sign will be biased. Consequently these corrections have a mean which takes the scaling form
\begin{equation}
    \delta \lambda^{\l w \r}_{n}= \sum_{m^{w} \neq n^{w}} \frac{ \lv\bra{n^{\l w \r}} P_{w'}\tilde{\rho}_{RQ}^{\l w' \r}\ket{m^{\l w\r}} \rv^{2}}{\lambda^{\l w\r}_{n} - \lambda^{\l w \r}_{m}}\sim \Omega\l w\r \frac{M^{\l w' \r}}{ \Delta\l w \r} = \frac{M^{\l w' \r}}{\delta\l w\r}.
\end{equation}
The sign of this mean flips between the top and bottom of the band to broaden it (level repulsion). Comparing this quantity to the bandwidth of $w$, $\Delta\l w \r$ we arrive at
\begin{equation}
    \frac{\delta \lambda^{\l w \r}_{n}}{\Delta\l w \r} = \frac{\Omega\l  w \r M^{\l w' \r}}{ \l \Delta\l w\r\r^{2}} = \Omega\l w \r \times \frac{1}{\Omega\l w'\r}\l\frac{P_{w'}}{q^{N+k}}\r^{2} \times \frac{q^{N+k}\Omega\l w \r}{P_{w}^{2}} = \frac{\Omega\l w' \r}{q^{N+k}} \l \frac{P_{w'} \Omega\l w \r}{P_{w} \Omega\l w' \r}\r^{2} =\frac{\Omega\l w' \r}{q^{N+k}} u^{2\l w' - w\r}.
\end{equation}
The factor $\Omega\l w' \r q^{ - \l N + k\r}$ suppresses this contribution so long as $w'$ is not the reservoir. If $w'$ is the reservoir, this term is still suppressed exponentially in the difference $w'-w$; however, when $w'$ is the reservoir and $w' -w \sim O(1)$, the corrections are allowed to be significant. The near hashing-weight bands are therefore allowed to experience substantial corrections (on the order of their own width) due to the reservoir.

While the above calculation captures the typical corrections, there are also corrections which are adversarial from the perspective of second-order perturbation theory, which is controlled by the ratio (which we call $g$) of the magnitude of the off-diagonal matrix element to the unperturbed level spacing. By selecting the $ \Omega^{\eta}\l w \r \equiv \l \Omega \l w \r\r^{\eta}$ eigenvalues nearest $\lambda_{n}$ (that is $\lv \lambda^{w}_{n} - \lambda^{w}_{m} \rv \sim \delta\l w\r \Omega^{\eta}\l w \r$) with $\eta <1$ we get
\begin{equation}
    g = \frac{\sqrt{M^{\l w'\r }}}{\Omega^{\eta}\l w \r\delta\l w\r} = \frac{P_{w'}}{P_{w}}\l\frac{\Omega^{3-2\eta}\l w \r}{q^{N+k}\Omega\l w'\r}\r^{1/2} = u^{w'-w}\l \frac{\Omega\l w' \r \Omega^{1-2\eta}\l w \r}{q^{N+k}}\r
\end{equation}
which is not always small when $\eta \leq 1/2$. When $g\gg 1 $ the corrrections can become larger than a straightforward perturbation theory approach would estimate. In this regime we can estimate the correction as
\begin{equation}
    \delta \lambda_{n}^{w}(m^{w})\approx \frac{{{\rm sgn}\l\lambda_{n}^{\l w \r} - \lambda_{m}^{\l w \r}\r}}{2}\sqrt{\l\lambda_{n}^{\l w \r} - \lambda_{m}^{\l w \r}\r^{2} + 4 M^{\l w'\r}} \sim\sqrt{M^{\l w' \r}}
\end{equation}
when $g \gtrsim 1 $. For all but very few isolated eigenvalues near the boundary and so long as the density of states has a convergent first order derivative, the signs are approximately unbiased (the biased part will contribute less than the fluctuations for $\eta < 2/3$) and the overall contribution will be central limiting. Therefore overall effect of these corrections can be estimated as
\begin{equation}
    \delta \lambda^{\l w\r }_{n} = \l \Omega^{\eta}\l w \r M^{w'}\r^{1/2} = \l \frac{\Omega^{\eta}\l w \r}{\Omega\l w' \r}\r^{1/2}\frac{P_{w'}}{q^{N+k}}.
\end{equation}
Comparing this to the width of band $w$ we get
\begin{equation}
    \frac{\delta \lambda^{\l w\r }_{n}}{\Delta\l w \r}\sim \frac{P_{w'}}{P_{w}} \l \frac{\Omega^{1+\eta}\l w \r}{q^{N+k}\Omega\l w' \r}\r^{1/2}   = u^{w'-w}\l \Omega^{\eta-1}\l w\r\frac{\Omega\l w'\r}{q^{N+k}}\r^{1/2}
\end{equation}
 which is manifestly small since $\eta \leq 1/2$ (when both $w'$ and $w$ are small, the factor of $q^{-\l N + k \r /2}$ makes this expression small and when $w'\gg w$ with $w\sim O(1)$ the $u^{w'-w}$ makes it small). Thus, we see that the second order corrections do not considerably broaden the bands.

\section{Soft postselection}\label{app:soft}
In this section, we expand upon the discussion of R\'enyi thresholds and postselection in~\autoref{sect:postselect} of the main text, deriving~\autoref{eq:renyi-wstar},~\autoref{eq:renyi}, and~\autoref{eq:reweight}.

Consider again the spectrum of $\tilde{\rho}_Q$. The contribution from weight $w$ errors to $\tilde{\rho}_Q^\alpha$ is, within the zeroth-order ansatz,
\begin{equation}
    \lambda(w)^\alpha q^k \Omega(w) = q^{k(1-\alpha)} q^{-N H(w/N)} (1-p)^{\alpha(N-w)} \left(\frac{p}{q^2-1}\right)^{\alpha w}.
\end{equation}
Maximizing this quantity over $w$ yields the expression for $w^*(p,\alpha)$, the dominant error weight, given in~\autoref{eq:renyi-wstar}. This in turn implies (cf.~\autoref{eq:renyi})
\begin{equation}\label{eq:renyi-sub}
S_\alpha(\tilde{\rho}_Q) \approx \frac{1}{1-\alpha} \glog_q \left[\lambda(w^*(p,\alpha)) q^k \Omega(w^*(p,\alpha))\right] = N H_\alpha(p) + k
\end{equation}
\autoref{eq:renyi-sub} breaks down once $S_\alpha(\tilde{\rho}_Q)$ exceeds the maximum allowed value of $N$.\footnote{One can interpret this as saying that a different saddle point takes over, which will become more clear in the Weingarten calculus analysis of the next section.} This implies a threshold $p_c^{(\alpha)}$ which satisfies
\begin{equation}
    H_\alpha\left(p_c^{(\alpha)}\right) = 1 - k/N. 
\end{equation}

Thus, $p_c^{(\alpha)}$ is the error rate at which the R\'enyi entropy $S_\alpha(\tilde{\rho}_Q)$, as well as the postselection acceptance probability $\tr{\tilde{\rho}_Q}^\alpha$, has a non-analyticity. By a similar argument, the R\'enyi entropy $S_\alpha(\tilde{\rho}_{RQ})$ is non-analytic at $H_\alpha(p) = 1 + k/N$ (independently verified using the Weingarten calculus in the next section), but this is a less physically meaningful transition with no corresponding interpretation for soft postselection, as we do not have access to the subsystem $R$.

Now let us consider the \textit{von Neumann} entropy of the reweighted density matrix $\tilde{\sigma}_{Q,\alpha}$. As a limit of Renyi-$n$ entropies, 
\begin{equation}\label{eq:vn-reweight}
S_{\rm vN}(\tilde{\sigma}_{Q,\alpha}) = \lim_{n\rightarrow 1} \frac{1}{1-n} \left[\glog_q\left({\sum_i\lambda_i^{n\alpha}}\right) - n \glog_q\left(\sum_i \lambda_i^\alpha\right)\right].
\end{equation}

Thus, the leading-$N$ behavior of $S_{\rm vN}(\tilde{\sigma}_{Q,\alpha})$ evaluates to
\begin{equation}
    S_{\rm vN}(\tilde{\sigma}_{Q,\alpha}) \approx \begin{cases}
    \lim_{n\rightarrow 1} \frac{1}{1-n}\left[(1-n\alpha)(N H_{n\alpha}(p) + k) - (1-\alpha)(N H_\alpha(p) + k)\right] & p < p_c^{(\alpha)} \\
    N & p > p_c^{(\alpha)}
    \end{cases}.
\end{equation}
The limit $n\rightarrow 1$ can be evaluated by brute force, but a more intuitive approach is to observe that within the zeroth order ansatz (which yields the correct leading-$N$ behavior up to the threshold $p_c^{(\alpha)}$), reweighting the eigenvalues directly corresponds to reweighting the binomial distribution of errors:
\begin{equation}\label{eq:reweight-lambda}
\lambda(w) \rightarrow \frac{\lambda(w)^\alpha}{\sum_w \lambda(w)^\alpha} \quad\Longleftrightarrow \quad 
p \rightarrow p_{\alpha} \equiv \frac{(q^2-1) \left(\frac{p}{q^2-1}\right)^\alpha}{(q^2-1) \left(\frac{p}{q^2-1}\right)^\alpha + (1-p)^\alpha} = \frac{w^*(p,\alpha)}{N}.
\end{equation}
We then recover that for $H_\alpha(p) < 1-k/N$ (cf.~\autoref{eq:vn-reweight}),
\begin{equation}
S_{\rm vN}(\tilde{\sigma}_{Q,\alpha}) \approx N H(w^*(p,\alpha)/N) + k.
\end{equation}
Since $H(w^*(p,\alpha)/N) < H_\alpha(p)$, $S_{\rm vN}(\tilde{\sigma}_{Q,\alpha})$ is \textit{discontinuous} as $N\rightarrow\infty$ at the transition. The finite-$N$ rounding of the transition is well approximated using the perturbative ansatz for the mean eigenvalues, as evidenced by the close agreement between the light red curve and the $N=13$ numerics in~\autoref{fig:postselect-renyi}(c).

In the main text, we further claimed that the \textit{coherent information} of the soft-postselected density matrix has a transition at the Renyi-$\alpha$ threshold. We do not have an ansatz for this quantity near the transition for finite $\alpha$, as the spectrum of
\begin{equation}
    \tilde{\sigma}_{RQ,\alpha} \equiv \frac{(\mathbbm{1}_R \otimes M_\alpha) \tilde{\rho}_{RQ} (\mathbbm{1}_R \otimes M_\alpha)}{\lambda_0^{1-\alpha}\tr{\tilde{\rho}_Q^\alpha}}
\end{equation}
requires a more detailed understanding of the overlap between eigenvectors of $\tilde{\rho}_Q$ and $\tilde{\rho}_{RQ}$. We can nevertheless argue for this transition as follows. For $p<p_c^{(\alpha)}$, reweighting $\lambda \rightarrow \lambda^\alpha$ effectively shifts the weight of a ``typical error'' from $pN$ (which, for $p>p_c$, is uncorrectable) to $w^*(p,\alpha)$. Since errors of this weight are correctable ($w^*(p,\alpha) < w_{\rm max}(p)$), the coherent information is maximal, in the limit $N\rightarrow \infty$. That is, to first approximation, each eigenvector in band $w=w^*(p,\alpha)$ of $\tilde{\rho}_{RQ}$ corresponds to $q^k$ degenerate eigenvectors in band $w=w^*(p,\alpha)$ of $\tilde{\rho}_Q$, so that $S_{\rm vN}(\tilde{\sigma})_{RQ,\alpha} = N H(w^*(p,\alpha)/N)$. This relationship breaks down above $p_c^{(\alpha)}$, as the $\alpha$-reweight fails to project into a well-defined band. 

In~\autoref{fig:postselect-renyi}(d), we obtained a good scaling collapse of the $\alpha=2$ soft-postselected coherent information with scaling argument $(p-p_c)N$ ($\nu=1$), in contrast with the $(p-p_c)N^{1/2}$ scaling ($\nu=2$) in the absence of postselection. Since the transition in the soft-postselected coherent information is governed by the same mechanism as the $\alpha$-R\'enyi entropy, we can heuristically argue for this scaling as follows. In~\autoref{app:macw} below, we will show that the R\'enyi $\alpha=2$ and the R\'enyi $\alpha=\infty$ (detection) transitions are determined by a common duality and share the same critical behavior. The detection threshold involves only one band (the $w=0$ band), so the width of the transition must be of only width $O(1)$ in number of errors, or $O(1/N)$ in $p$, which is the narrowest possible scaling. We further expect, and calculations with the perturbative ansatz support, that $\nu$ is pinned to 1 in the entire range $\alpha \in [2,\infty)$. We leave open the trend of $\nu$ in the interval $\alpha \in (1,2)$.

\section{Hashing Threshold and R\'enyi Thresholds via Weingarten Calculus}\label{app:weingarten}

In this Appendix, we provide a direct analytical derivation of the hashing threshold and higher R\'enyi thresholds in Haar-random encodings, in the limit of large system size $N \to \infty$. This calculation provides an interesting alternative derivation of the hashing bound \cite{Bennett1996hashing,DiVincenzo1997}, a lower bound on the quantum channel capacity of Pauli decoherence channels, which is typically proven using random stabilizer encodings and counting arguments \cite{wildeQuantumInformationTheory2013,preskillLectureNotesPhysics2018}.

The derivation employs standard techniques of Weingarten calculus \cite{collins_integration_2006,Mele2024introductiontohaar} to compute the Haar averages of the purities $\ztr \tilde{\rho}_{Q}^{n}$ and $\ztr \tilde{\rho}_{RQ}^{n}$ of the decohered density matrices $\tilde{\rho}_{Q}$ and $\tilde{\rho}_{RQ}$. These can then be leveraged to obtain the average von Neumann entropies $S_{\vn}(\tilde{\rho}_{Q})$ and $S_{\vn}(\tilde{\rho}_{RQ})$ via the replica limit
\begin{equation}
\label{eq:vN_replica_limit}
    \haar S_{\vn}(\rho) = \lim_{n \to 1} \tilde{S}_{n}(\rho), \quad \tilde{S}_{n}(\rho) \equiv \frac{1}{1-n} \zlog_{q} \qty{ \haar \ztr\rho^{n} } ,
\end{equation}
where $\haar$ denotes the average over Haar-random encoding unitaries $U$. The quantity $\tilde{S}_{n}(\rho)$ can be regarded as an ``annealed average'' R\'enyi entropy, and we shall later find that its leading-in-$N$ behavior is identical to the ``quenched average'' R\'enyi entropy $\haar S_{n}(\rho)$ in the present context.

The calculation of average purities $\haar \ztr \tilde{\rho}_{Q}^{n}$ and $\haar \ztr \tilde{\rho}_{RQ}^{n}$ will be facilitated by the following mathematical identity~\cite{collins_integration_2006}: for an ensemble of $D \times D$ random unitary matrices $U$ distributed by the Haar measure, the average of the $n$th \textit{moment operator} $(U \otimes U^{*})^{\otimes n}$ is given by
\begin{equation}
    \label{eq:moment_op}
    \haar (U \otimes U^{*})^{\otimes n} = \sum_{\sigma, \tau \in \mathcal{P}_{n}} \Wg{n}{D}(\sigma^{-1} \tau) \kket{\sigma} \bbra{\tau} ,
\end{equation}
where $\sigma$ and $\tau$ are elements of the $n$-fold permutation group $\mathcal{P}_{n}$, and the states $\kket{\sigma}$ for each permutation $\sigma \in \mathcal{P}_{n}$ are defined by the matrix elements
\begin{equation}
    \bbra{a_{1} \bar{a}_{1} \ldots a_{n} \bar{a}_{n}} \sigma \rangle \! \rangle = \delta_{a_{\sigma(1)} \bar{a}_{1}} \ldots \delta_{a_{\sigma(n)} \bar{a}_{n}} .
\end{equation}
In the following, we shall regard each $a_{i}$ as a ``ket'' index and each $\bar{a}_{i}$ as a ``bra'' index, so that the state $\kket{\sigma}$ contracts each $i$th bra index with the $\sigma(i)$th ket index. Finally, $\Wg{n}{D}$ is the Weingarten function \cite{weingarten_asymptotic_1978,collins_integration_2006}, which is a class function on $\mathcal{P}_{n}$. Although it has an exact (albeit complicated) closed-form expression \cite{collins_integration_2006}, we shall only need the asymptotic scaling \cite{weingarten_asymptotic_1978} $\Wg{n}{D}(\sigma) = \frac{1}{D^{n}} \qty[ \delta_{\sigma, \mathbb{I}} + \mathcal{O}(D^{-1}) ]$ in what follows, where $\mathbb{I}$ is the identity permutation.

To compute the purities $\ztr \rho_{X}^{n}$ for the subsystem $X$ using the identity \eqref{eq:moment_op}, it is convenient to employ a vectorized notation \cite{Mele2024introductiontohaar}. We first represent $\rho_{RQ} \mapsto \kket{\rho_{RQ}}$ as a state in a doubled Hilbert space, given explicitly by
\begin{equation}
    \kket{\rho_{RQ}} = \Big[ (\mathbbm{1}_{R} \otimes U_{Q}) \otimes (\mathbbm{1}_{\bar{R}} \otimes U^{*}_{\bar{Q}}) \Big] \ket{\Phi_{RL}}  \ket{0}_{A} \ket{\Phi^{*}_{\bar{R} \bar{L}}} \ket{0^{*}_{\bar{A}}}
\end{equation}
where $\bar{R}$, $\bar{Q}$, $\bar{L}$, and $\bar{A}$ are the analogs of subsystems $R$, $Q$, $L$, and $A$ in the ``bra'' Hilbert space, and $*$ denotes complex conjugation in the computational basis. In this vectorized Hilbert space, the depolarizing channel $\mathcal{N}_{p}$ acts as a nonunitary operator which couples the ``ket'' and ``bra'' Hilbert spaces: specifically, on each site $i \in Q$, the depolarizing channel $\mathcal{N}_{i,p}$ acts as the vectorized operator
\begin{equation}\label{eq:depol-gamma}
    \begin{split}
        \hat{\mathcal{N}}_{i,p} &= (1-p) \mathbbm{1} + \frac{p}{q^{2} - 1} \sum_{\mu = 1}^{q^{2} - 1} E^{\mu}_{i} \otimes E^{\mu *}_{i} \\
        &\equiv \sum_{\mu = 0}^{q^{2} - 1} p_{\mu} E^{\mu}_{i} \otimes E^{\mu *}_{i},
    \end{split}
\end{equation}
where we have defined $p_{0} = 1-p$ and $p_{\mu} = p/(q^{2} - 1)$ for $\mu = 1, \ldots , q^{2} - 1$. The total vectorized channel is $\hat{\mathcal{N}}_{p} = \bigotimes_{i \in Q} \hat{\mathcal{N}}_{i, p}$, and the decohered density matrix is $\kket{\tilde{\rho}_{RQ}} = \hat{\mathcal{N}}_{p} \kket{\rho_{RQ}}$. 

\begin{figure}[t]
    \centering
    \includegraphics[width=\textwidth]{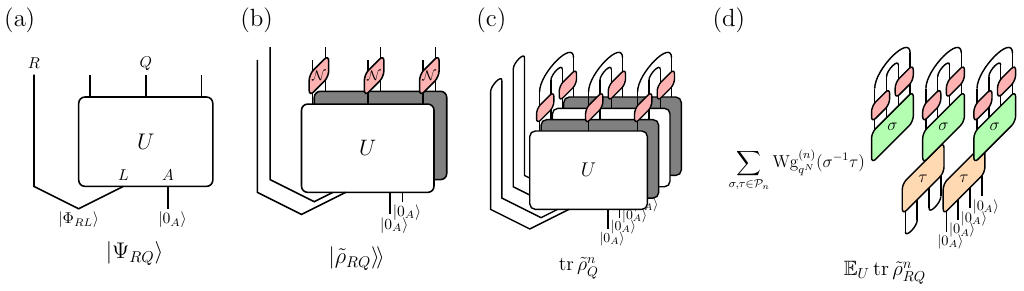}
    \caption{Tensor network diagrams for the sequence of transformations used to compute the averaged purity $\haar \ztr \tilde{\rho}_Q^n$. The corresponding diagram for $\haar \ztr \tilde{\rho}_Q^n$ is similar. (a) The initial encoded state $\ket{\Psi_{RQ}}$, consisting of $k$ logical qudits $L$ maximally entangled with a $k$-qudit reference $R$ and an auxiliary $N-k$ qudits $A$ initialized in a pure state $\ket{0_A}$, encoded using a Haar-random unitary $U$. (b) In vectorized notation, each local decoherence channel $\mathcal{N}_{i,p}$ acts as a nonunitary operator which couples the ``ket'' and ``bra'' copies. The dark gray box is $U^*$ acting on the ``bra'' copy. (c) The $n$th purity depicted using vectorized notation, with $n = 2$ for simplicity. The partial trace over $R$ is implemented by contracting with the $\bbra{\mathbb{I}}$ permutation, while the multiplication of adjacent density matrices is implemented by contracting with the $\bra{\mathbb{C}}$ permutation. (d) After averaging over unitary gates, the resulting tensor network can be easily evaluated for the dominant permutations $\sigma = \tau = \mathbb{I}, \mathbb{C}$.}
    \label{fig:hashing_weingarten}
\end{figure}

To represent $\ztr \tilde{\rho}_{X}^{n}$ in vectorized notation, we further replicate $\kket{\tilde{\rho}_{RQ}}$ $n$ times to obtain $\kket{\tilde{\rho}_{RQ}^{\otimes n}}$, and implement the partial trace and contraction of adjacent density matrices using the permutation states $\kket{\sigma}$ previously defined\footnote{In un-vectorized notation, this equation is simply $\ztr \tilde{\rho}_{X}^{n} = \ztr [ \hat{\mathbb{C}}_{X} \tilde{\rho}_{RQ}^{\otimes n} ]$, where $\hat{\mathbb{C}}_{X}$ is a cyclic permutation operator acting on subsystem $X$. This is a straightforward generalization of the more familiar equation for the case $n = 2$, where $\hat{\mathbb{C}}_{X}$ is simply a SWAP operation on $X$.}:
\begin{equation}
    \ztr \tilde{\rho}_{X}^{n} = \bbra{\mathbb{C}_{X}, \mathbb{I}_{X^{c}}} \tilde{\rho}^{\otimes n}_{RQ} \rangle \! \rangle , \quad \bbra{\mathbb{C}_{X}, \mathbb{I}_{X^{c}}} \equiv \bigotimes_{i \in X} \bbra{\mathbb{C}_{i}} \bigotimes_{i \in X^{c}} \bbra{\mathbb{I}_{i}} ,
\end{equation}
where $\mathbb{C} = (1 \ \ldots \ n)$ is the cyclic permutation, and $X^{c}$ is the complement of $X$. As a final step, since subsystem $R$ is maximally entangled with the initial state of $Q$, we can represent its contraction with a permutation $\sigma$ at the end of the circuit by a modified initial state of $L$. Specifically, the identity\footnote{In this equation, we have assumed $\ket{\Phi_{RL}} = \frac{1}{\sqrt{q^{k}}} \sum_{\vb{a}} \ket{\vb{a}_{R}} \ket{\vb{a}_{L}}$ is the standard Bell state, where $\ket{\vb{a}_R}$ is a complete orthonormal basis for $R$ (and similarly for $L$). Any other maximally entangled state is related to $\ket{\Phi_{RL}}$ by a unitary performed on $L$ alone, which can subsequently be absorbed into the random unitary $U$.} $\bbra{\sigma_{R}} \Phi_{RL}^{\otimes 2n} \rangle \! \rangle = \frac{1}{q^{nk}} \kket{\sigma_{L}}$ allows us to finally express the $n$th purities as
\begin{equation}
    \ztr \tilde{\rho}_{Q}^{n} = \frac{1}{q^{nk}} \bbra{\mathbb{C}_{Q}} \hat{\mathcal{N}}_{p}^{\otimes n} (U \otimes U^{*})^{\otimes n} \kket{\mathbb{I}_{L}, 0_{A}}, \quad \ztr \tilde{\rho}_{RQ}^{n} = \frac{1}{q^{nk}} \bbra{\mathbb{C}_{Q}} \hat{\mathcal{N}}_{p}^{\otimes n} (U \otimes U^{*})^{\otimes n} \kket{\mathbb{C}_{L}, 0_{A}} .
\end{equation}
This sequence of rewritings is depicted diagrammatically in Fig.~\ref{fig:hashing_weingarten}. Altogether, we have represented $\ztr \tilde{\rho}_{Q}^{n}$ as a particular matrix element of the moment operator $(U \otimes U^{*})^{\otimes n}$. 

Employing the identity \eqref{eq:moment_op} with $D = q^{N}$, we obtain the result
\begin{equation}
    \label{eq:avg_rhoQ}
    \haar \ztr \tilde{\rho}_{Q}^{n} = \frac{1}{q^{nk}} \sum_{\sigma, \tau \in \mathcal{P}_{n}} \Wg{n}{q^{N}}(\sigma^{-1} \tau) \bbra{\mathbb{C}_{Q}} \hat{\mathcal{N}}_{p}^{\otimes n} \kket{\sigma_{Q}} \bbra{\tau_{Q}} \mathbb{I}_{L}, 0_{A} \rangle \! \rangle ,
\end{equation}
and similarly for $\ztr \tilde{\rho}_{RQ}^{n}$. Each of the matrix elements above factorizes across sites in $Q$: the second matrix element in the above is simply given by
\begin{equation}
    \bbra{\tau_{Q}} \mathbb{I}_{L}, 0_{A} \rangle \! \rangle = \prod_{i \in L} \bbra{\tau_{i}} \mathbb{I}_{i} \rangle \! \rangle \prod_{i \in A} \bbra{\tau_{i}} 0_{i} \rangle \! \rangle = q^{k(n - \abs{\tau})} , 
\end{equation}
where we have used the identities $\bbra{\sigma_{i}} \tau_{i} \rangle \! \rangle = q^{n - \abs{\sigma^{-1} \tau}}$ and $\bbra{\tau_{i}} 0_{i} \rangle \! \rangle = 1$, where $\abs{\tau}$ is the number of transpositions required to construct $\tau$ from the identity permutation. Similarly, the first matrix element is given by
\begin{equation}
    \bbra{\mathbb{C}_{Q}} \hat{\mathcal{N}}_{p}^{\otimes n} \kket{\sigma_{Q}} = \prod_{i \in Q} \bbra{\mathbb{C}_{i}} \hat{\mathcal{N}}_{i,p} \kket{\sigma_{i}} = \qty[ \sum_{\mu_{1}, \ldots , \mu_{n} = 0}^{q^{2} - 1} p_{\mu_{1}} \ldots p_{\mu_{n}} \bbra{\mathbb{C}_{i}} E^{\mu_{1}}_{i} \otimes E^{\mu_{1}*}_{i} \otimes \ldots \otimes E^{\mu_{n}}_{i} \otimes E^{\mu_{n}*}_{i} \kket{\sigma_{i}} ]^{N} .
\end{equation}

Thus far, we have an exact but intractable result for the average purities. An enormous simplification arises in the limit $N \to \infty$, where a single term in the sum \eqref{eq:avg_rhoQ} exponentially dominates all other terms. In this limit, we can obtain the leading-$N$ behavior of $\tilde{S}_{n}(\tilde{\rho}_{Q})$ and $\tilde{S}_{n}(\tilde{\rho}_{RQ})$ by keeping only the single dominant term in the sum. Specifically, it is readily observed that the summand in Eq.~\eqref{eq:avg_rhoQ} is maximized either by setting $\sigma = \tau = \mathbb{I}$ or $\sigma = \tau = \mathbb{C}$, depending on the value of $p$. This substitution significantly simplifies the computation of the matrix elements $\bbra{\mathbb{C}_{i}} \hat{\mathcal{N}}_{i,p} \kket{\sigma_{i}}$: when $\sigma = \mathbb{I}$, we immediately obtain $\bbra{\mathbb{C}_{i}} \hat{\mathcal{N}}_{i,p} \kket{\mathbb{I}_{i}} = \bbra{\mathbb{C}_{i}} \mathbb{I}_{i} \rangle \! \rangle = q$, since $\mathcal{N}_{i,p}$ is a unital channel; and when $\sigma = \mathbb{C}$, we can evaluate the sum in brackets above as follows:
\begin{equation}
    \begin{split}
        \sum_{\mu_{1}, \ldots , \mu_{n} = 0}^{q^{2} - 1} p_{\mu_{1}} \ldots p_{\mu_{n}} & \bbra{\mathbb{C}_{i}} E^{\mu_{1}}_{i} \otimes E^{\mu_{1}*}_{i} \otimes \ldots \otimes E^{\mu_{n}}_{i} \otimes E^{\mu_{n}*}_{i} \kket{\mathbb{C}_{i}} \\
        &= \sum_{\mu_{1}, \ldots , \mu_{n} = 0}^{q^{2} - 1} p_{\mu_{1}} \ldots p_{\mu_{n}} \ztr[E^{\mu_{1}}_{i} E^{\mu_{2}}_{i}] \ztr[E^{\mu_{2}}_{i} E^{\mu_{3}}_{i}] \ldots \ztr[E^{\mu_{n}}_{i} E^{\mu_{1}}_{i}] \\
        &= \sum_{\mu_{1}, \ldots , \mu_{n} = 0}^{q^{2} - 1} p_{\mu_{1}} \ldots p_{\mu_{n}} (q \delta_{\mu_{1} \mu_{2}}) (q \delta_{\mu_{2} \mu_{3}}) \ldots (q \delta_{\mu_{n} \mu_{1}}) \\
        &= q^{n} \sum_{\mu = 0}^{q^{2} - 1} p_{\mu}^{n} \\
        &= q^{n} q^{-(n-1) H_{n}(p)} ,
    \end{split}
\end{equation}
where the $n$th R\'enyi-Shannon entropy $H_n(p)$ is defined in Eq.~\eqref{eq:shannon-renyi}. Note in particular that $H(p) \equiv -\sum_{\mu = 0}^{q^{2} - 1} p_{\mu} \zlog\, p_{\mu} = \lim_{n \to 1} H_{n}(p)$. Using this result, along with $\Wg{n}{q^{N}}(\mathbb{I}) = q^{-nN}$, we obtain the following final expressions for the averaged purities:
\begin{equation}
    \begin{split}
        \haar \ztr \tilde{\rho}_{Q}^{n} &= \begin{cases}
        q^{-(n-1)[ k + N H_{n}(p) ]}, & H_{n}(p) \leq 1 - \frac{k}{N}, \\
        q^{-(n-1)N}, & H_{n}(p) \geq 1 - \frac{k}{N}
    \end{cases}, \\
    \haar \ztr \tilde{\rho}_{RQ}^{n} &= \begin{cases}
        q^{-(n-1) N H_{n}(p)}, & H_{n}(p) \leq 1 + \frac{k}{N}, \\
        q^{-(n-1)(N+k)}, & H_{n}(p) \geq 1 + \frac{k}{N}
    \end{cases} .
    \end{split}
\end{equation}
In both purities, the first case corresponds to setting $\sigma = \tau = \mathbb{C}$ in the sum, while the second case corresponds to setting $\sigma = \tau = \mathbb{I}$. From these expressions, we obtain our final result for the averaged von Neumann entropies:
\begin{equation}
 \begin{split}
        \haar S_{\vn}(\tilde{\rho}_{Q}) &= \begin{cases}
        k + N H(p), & H(p) \leq 1 - \frac{k}{N} \\
        N, & H(p) \geq 1 - \frac{k}{N}
    \end{cases}, \\
    \haar S_{\vn}(\tilde{\rho}_{RQ}) &= \begin{cases}
        N H(p), & H(p) \leq 1 + \frac{k}{N} \\
        N + k, & H(p) \geq 1 + \frac{k}{N}
    \end{cases} ,
 \end{split}
\end{equation}
which is precisely the behavior predicted in the main text. The average coherent information $\haar I_{c}(R \rangle Q)$ therefore obeys precisely the ansatz of Eq.~\eqref{eq:Ic_fixedweight_ansatz} of the main text.

The preceding calculation can easily be extended to compute the average R\'enyi entropies, $\haar S_{\alpha}(\tilde{\rho}_{Q})$ and $\haar S_{\alpha}(\tilde{\rho}_{RQ})$. This is facilitated with a modified replica limit:
\begin{equation}
    \haar S_{\alpha}(\rho) = \lim_{\ell \to 0} \tilde{S}_{\alpha,\ell}(\rho), \quad \tilde{S}_{\alpha, \ell}(\rho) = \frac{1}{\ell(1-\alpha)} \zlog \qty{ \haar \qty[ \ztr \rho^{\alpha} ]^{\ell} } .
\end{equation}
We are thus left to compute the Haar averages of the $\ell$th moments of the $\alpha$th purities, $\haar [\ztr \tilde{\rho}_{Q}^{\alpha}]^{\ell}$ and $\haar [\ztr \tilde{\rho}_{RQ}^{\alpha}]^{\ell}$. The calculation is nearly identical to the previous one (which corresponds to $\ell = 1$), so we omit the full calculation and simply mention the differences. First, similar to the previous case, we can express $[\ztr \rho_{X}^{\alpha}]^{\ell} = \bbra{\mathbb{D}_{X}, \mathbb{I}_{X^{c}}} \rho_{RQ}^{\otimes n} \rangle \! \rangle$ using a vectorized and $n$-fold replicated density matrix $\kket{\rho^{\otimes n}_{RQ}}$, where $n = \alpha \ell$, and the cyclic permutation state in region $X$ is replaced with the modified cyclic permutation
\begin{equation}
    \mathbb{D} = (1 \ \ldots \ \alpha) ((\alpha+1) \ \ldots \ 2\alpha) \ldots ((\alpha\ell - \alpha) \ \ldots \ \alpha\ell) .
\end{equation}
Correspondingly, the dominant permutations in the Weingarten sum are $\sigma = \tau = \mathbb{D}$ and $\sigma = \tau = \mathbb{I}$. A calculation of $\bbra{\mathbb{D}_{i}} \hat{\mathcal{N}}_{i, p} \kket{\mathbb{D}_i}$ proceeds identically as before, with the result
\begin{equation}
    \bbra{\mathbb{D}_{i}} \hat{\mathcal{N}}_{i, p} \kket{\mathbb{D}_i} = \qty[ q^{\alpha} q^{-(\alpha-1) H_{\alpha}(p)} ]^{\ell} .
\end{equation}
Using this result, one finds that the leading-$N$ behavior of $\haar [\ztr \tilde{\rho}_{Q}^{\alpha}]^{\ell}$ factorizes across each of the $\ell$ copies of the purity, i.e.,
\begin{equation}
    \haar [\ztr \tilde{\rho}_{Q}^{\alpha}]^{\ell} = [ \haar \ztr \tilde{\rho}_{Q}^{\alpha}]^{\ell} ,
\end{equation}
and similarly for $\tilde{\rho}_{RQ}$. This immediately implies that each $\tilde{S}_{\alpha,\ell}$ is independent of $\ell$ and equal to the ``annealed'' R\'enyi entropy $\tilde{S}_{\alpha}$ previously computed. Therefore, the average R\'enyi entropies are simply given by
\begin{equation}
     \begin{split}
        \haar S_{\alpha}(\tilde{\rho}_{Q}) &= \begin{cases}
        k + N H_{\alpha}(p), & H_{\alpha}(p) \leq 1 - \frac{k}{N} \\
        N, & H_{\alpha}(p) \geq 1 - \frac{k}{N}
    \end{cases}, \\
    \haar S_{\alpha}(\tilde{\rho}_{RQ}) &= \begin{cases}
        N H_{\alpha}(p), & H_{\alpha}(p) \leq 1 + \frac{k}{N} \\
        N + k, & H_{\alpha}(p) \geq 1 + \frac{k}{N}
    \end{cases} ,
 \end{split}
\end{equation}
again in agreement with the main text [\autoref{eq:renyi}].
\color{black}
\section{Heralded Erasure Errors}\label{app:erasures}
In the main text, we have focused on an error model consisting of i.i.d. depolarization channels on each qudit. A natural generalization is to consider error models with additional structure. Here we consider the case of \emph{heralded} errors, where the observer has detailed knowledge of which qudits are affected or unaffected by decoherence. 

As a simple model of heralded errors, we consider the erasure channel $\mathcal{E}_{i,p}$ which erases the $i$th qudit with probability $p$:
\begin{equation}
    \mathcal{E}_{i, p}(\rho) = (1 - p) \rho \otimes \dyad{0}_{i_M} + p \, \ztr_i[\rho] \otimes \frac{\mathbbm{1}_i}{q} \otimes \dyad{1}_{i_M} .
\end{equation}
The total channel is then $\mathcal{E}_p \equiv \bigotimes_{i \in Q} \mathcal{E}_{i,p}$. Note that we have introduced a flag qubit $i_M$ for each $i$th qudit, which tracks whether a qudit has been erased. The set of all such flag qubits constitutes another auxiliary subsystem $M$, which is classically correlated with the density matrix on $RQ$, and provides the `observer' with classical information on the locations of erasures. The full density matrix is given by
\begin{equation}
    \tilde{\rho}_{RQM} = \sum_{\vb{m}} p_{\vb{m}} \rho_{RQ}(\vb{m}) \otimes \dyad{\vb{m}}_M ,
\end{equation}
where $\vb{m} \equiv \qty{m_i}_{i \in Q}$ indicates a trajectory of erasures, with $m_i = 1$ ($m_i = 0$) indicating the presence (absence) of an erasure; $p_{\vb{m}} \equiv (1 - p)^{N - \abs{\vb{m}}} p^{\abs{\vb{m}}}$ is the total probability of the trajectory $\vb{m}$; and $\rho_{RQ}(\vb{m})$ is the density matrix obtained from maximally depolarizing each qudit on which $m_i = 1$. Note that tracing over $M$ converts the channel $\mathcal{E}_p$ back to a depolarization channel (see~\autoref{app:dual} of this Supplement).

We are interested in the coherent information $I_c(R \rangle QM)$, which we can equivalently express as
\begin{equation}
    I_c(R \rangle QM) = \sum_{\vb{m}} p_{\vb{m}} I_c(R \rangle Q | \vb{m}) ,
\end{equation}
where $I_c(R \rangle Q | \vb{m}) \equiv S_{\vn}[\rho_Q(\vb{m})] - S_{\vn}[\rho_{RQ}(\vb{m})]$ is the \emph{conditional} coherent information, given the erasure trajectory $\vb{m}$. Upon averaging over Haar-random encodings, the average conditional coherent information $\haar I_c(R \rangle Q | \vb{m})$ depends only on the total number $m = \abs{\vb{m}}$ of erasures and not on their locations. We therefore have
\begin{equation}
\label{eq:Ic_erasure_sum}
    \haar I_c(R \rangle QM) = \sum_{m = 0}^N \binom{N}{m} (1 - p)^{N-m} p^m I_c(m) ,
\end{equation}
where $I_c(m)$ is the average coherent information upon erasing $m$ qudits from $Q$. In the limit $N \to \infty$, we can evaluate this sum using the saddle-point method, so that $\haar I_c(R \rangle QM) \approx I_c(pN)$. In what follows, we will first evaluate $I_c(m)$ with a fixed number of easures, and then we will discuss the full average expression in more detail.

To determine the leading behavior of $I_c(m)$ as a function of $m$, a useful heuristic is to regard the full encoded state $\ket{\Psi_{RQ}}$ as a Haar-random state on all of $RQ$. Let us first divide $Q$ into an $m$-qudit subsystem $E$ of erased qudits and an $(N-m)$-qudit system $K$ of unaffected qubits; due to the Haar-random encoding, the choice of subsystems is immaterial. We can then obtain the leading contributions to $S_{\vn}(\rho_K)$ and $S_{\vn}(\rho_{RK})$ via Page's theorem \cite{Page1993}. There are two obvious cases to consider:
\begin{enumerate}
    \item If $RE$ comprises less than half of the total number of qubits (i.e., $k + m < \frac{1}{2}(k + N)$), then the reduced density matrix $\rho_{RE}$ is maximally mixed, and in particular $\rho_{RE} = \rho_R \otimes \rho_E$. Since $RKE$ is globally pure, this implies that $R$ is maximally entangled with $K$, and so $I_c(m) = k$.
    \item Conversely, if $RK$ comprises less than half of the total number of qubits (i.e., $k + (N - m) < \frac{1}{2} (k + N)$), then $\rho_{RK}$ is maximally mixed. This gives $I_c(m) = - k$.
\end{enumerate}
If we assume that $I_c(m)$ linearly interpolates between these two extremes, then we obtain the ansatz
\begin{equation}
\label{eq:coherent_info_erasure_ansatz}
    \haar I_c(m) = \begin{cases}
        k, & m < \frac{1}{2}(N - k) \\
        N - 2m, & \frac{1}{2} (N - k) \leq m \leq \frac{1}{2}(N+k) \\
        -k, & \frac{1}{2}(N + k) < m
    \end{cases} .
\end{equation}
Assuming $m = pN$, this simple-minded ansatz immediately predicts that $\haar I_c(R \rangle QM)$ is maximal so long as $k/N < 1 - 2p$. Our ansatz therefore predicts that the Haar-random encoding saturates the well-known quantum channel capacity of the erasure channel \cite{wildeQuantumInformationTheory2013}, as might have been expected.

We can obtain the same result more systematically by employing the Weingarten calculus in the same manner as Appendix~\ref{app:weingarten}. Employing identical manipulations as in the case of depolarization, we arrive at the analogous formulae for the $n$th purities:
\begin{equation}
    \ztr \rho_K^n = \frac{1}{q^{nk}} \bbra{\mathbb{C}_K, \mathbb{I}_E} (U \otimes U^*)^{\otimes n} \kket{\mathbb{I}_L, 0_A}, \quad \ztr \rho_{RK}^n = \frac{1}{q^{nk}} \bbra{\mathbb{C}_K, \mathbb{I}_E} (U \otimes U^*)^{\otimes n} \kket{\mathbb{C}_L, 0_A} ,
\end{equation}
where the contraction with $\mathbb{I}$ on subsystem $E$ arises due to the erasure channel. Averaging over random encodings, we obtain the average purity
\begin{equation}
    \begin{split}
        \haar \ztr \rho_K^n &= \frac{1}{q^{nk}} \sum_{\sigma, \tau \in \mathcal{P}_n} \Wg{n}{q^N}(\sigma^{-1} \tau) \bbra{\mathbb{C}_K, \mathbb{I}_E} \sigma_Q \rangle \! \rangle \langle \! \langle \tau_Q \kket{\mathbb{I}_L, 0_A}  \\
        &= \frac{1}{q^{nk}} \sum_{\sigma, \tau \in \mathcal{P}_n} \Wg{n}{q^N}(\sigma^{-1} \tau) q^{(N-m)(n - \abs{\mathbb{C}^{-1} \sigma})} q^{m(n - \abs{\sigma})} q^{k(n - \abs{\tau})} , \\
    \end{split}
\end{equation}
and similarly for $\haar \ztr \rho_{RK}^n$.

We now once again work in the limit $N \to \infty$, where a single term in the above sum exponentially dominates. The leading behavior is obtained by setting $\sigma = \tau = \mathbb{C}$ when $N - m > m + k$, and by setting $\sigma = \tau = \mathbb{I}$ when $N - m < m + k$. This immediately gives
\begin{equation}
    \haar \ztr \rho_K^n = \begin{cases}
        q^{-(n - 1)(m+k)}, & m < \frac{1}{2}(N-k) \\
        q^{-(n - 1)(N - m)}, & m \geq \frac{1}{2}(N - k)
    \end{cases} .
\end{equation}
An identical calculation for $\haar \ztr \rho_{RK}^n$ yields
\begin{equation}
    \haar \ztr \rho_{RK}^n = \begin{cases}
        q^{-(n-1) m}, & m \leq \frac{1}{2} (N + k) \\
        q^{-(n-1)(N - m + k)}, & m > \frac{1}{2}(N + k)
    \end{cases} .
\end{equation}
Taking the replica limit $n \to 1$ using Eq.~\eqref{eq:vN_replica_limit}, we obtain the von Neumann entropies
\begin{equation}
    \haar S_{\vn}(\rho_K) = \begin{cases}
        m + k, & m < \frac{1}{2} (N - k) \\
        N - m, & m \geq \frac{1}{2} (N - k)
    \end{cases}, \quad \haar S_{\vn}(\rho_{RK}) = \begin{cases}
        m, & m \leq \frac{1}{2} (N + k) , \\
        N - m + k, & m > \frac{1}{2} (N + k)
    \end{cases} .
\end{equation}
The difference of these two expressions immediately yields $I_c(m) = \haar [S_{\vn}(\rho_K) - S_{\vn}(\rho_{RK})]$ given by Eq.~\eqref{eq:coherent_info_erasure_ansatz}.

A benefit of this systematic approach is that we can derive subleading corrections to the coherent information by determining the next leading permutations in the Weingarten sums. Considering $\haar \ztr \rho_K^n$ first, the largest term in the sum for $m < \frac{1}{2}(N -k)$ is when $\sigma = \tau = \mathbb{C}$, and the next largest terms are when $\sigma = \tau = \mathbb{C} \circ \mathbb{T}$, where $\mathbb{T}$ is a single transposition. There are $\binom{n}{2}$ possible choices for $\mathbb{T}$. On the other hand, the largest term for $m \geq \frac{1}{2}(N - k)$ is when $\sigma = \tau = \mathbb{I}$, and the next largest are when $\sigma = \tau = \mathbb{T}$ is a single transposition. Once again, there are $\binom{n}{2}$ possible choices. We therefore obtain the result
\begin{equation}
    \haar \ztr \rho_K^n = \begin{cases}
        q^{(n - 1)(m + k)} \qty[ 1 + \binom{n}{2} q^{-(N-k-2m)} ], & m < \frac{1}{2}(N - k) \\
        q^{(n - 1)(N - m)} \qty[ 1 + \binom{n}{2} q^{-(2m - N + k)} ], & m \geq \frac{1}{2}(N - k)
    \end{cases} .
\end{equation}
Note that these subleading corrections are exponentially small in $\abs{m - \frac{1}{2}(N-k)}$. An identical analysis holds for $\haar \ztr \rho_{RK}^n$, and gives the result
\begin{equation}
    \haar \ztr \rho_{RK}^n = \begin{cases}
        q^{-(n-1)m} \qty[ 1 + \binom{n}{2} q^{-(N - 2m + k)} ] , & m \leq \frac{1}{2}(N + k) , \\
        q^{-(n-1)(N - m + k)} \qty[ 1 + \binom{n}{2} q^{-(2m - N - k)} ], & m > \frac{1}{2}(N + k)
    \end{cases} ,
\end{equation}
where the subleading corrections are now exponentially small in $\abs{m - \frac{1}{2}(N + k)}$. Altogether, we obtain the replica limits
\begin{equation}
    \begin{split}
        \haar S_{\vn}(\rho_K) &= \begin{cases}
            m + k - \frac{1}{2 \ln q} q^{-(N - k - 2m)}, & m < \frac{1}{2}(N - k) \\
            N - m - \frac{1}{2 \ln q} q^{-(2m - N + k)}, & m \geq \frac{1}{2}(N - k)
        \end{cases}, \\
        \haar S_{\vn}(\rho_{RK}) &= \begin{cases}
            m - \frac{1}{2 \ln q} q^{-(N - 2m + k)}, & m \leq  \frac{1}{2}(N + k) \\
            N - m + k - \frac{1}{2 \ln q} q^{-(2m - N - k)}, & m > \frac{1}{2}(N + k)
        \end{cases} .
    \end{split}
\end{equation}
We finally obtain $I_c(m)$ from the difference of the above two expressions. Note that the subleading corrections to the ansatz \eqref{eq:coherent_info_erasure_ansatz} are precisely the usual corrections to Page's theorem \cite{Page1993}, which could have been immediately guessed from our original ansatz. These corrections are non-negligible only in an $\mathcal{O}(1)$ window of $m$ values centered on $m = \frac{1}{2}(N \pm k)$.  As a function of $p$, $I_c(pN)$ therefore obeys the ansatz \eqref{eq:coherent_info_erasure_ansatz} everywhere besides an $\mathcal{O}(1/N)$ window centered on $p = \frac{1}{2}(1 \pm k/N)$.

\begin{figure}
    \centering
    \includegraphics[width=\textwidth]{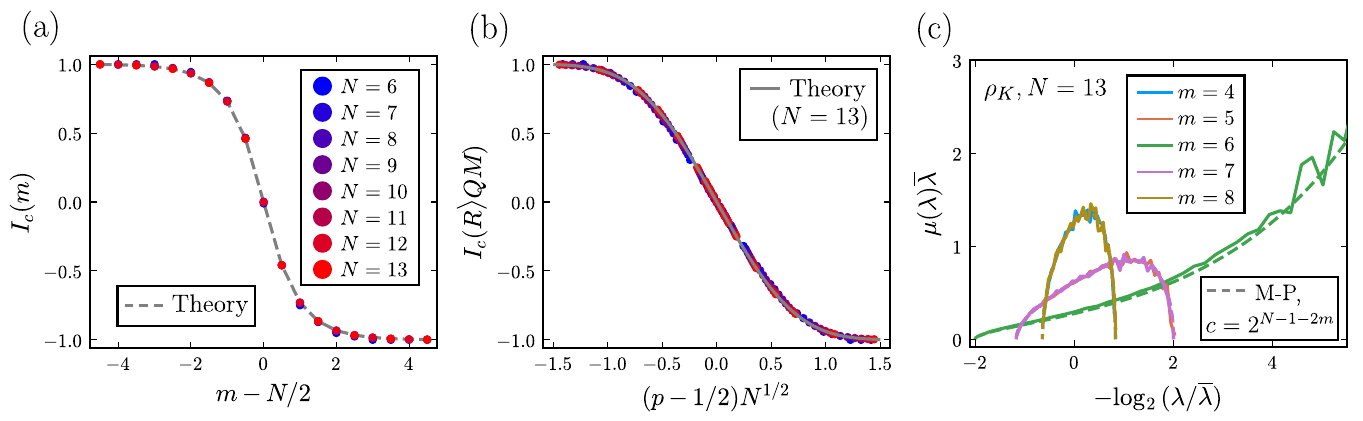}
    \caption{Coherent information under heralded erasure errors. (a) Coherent information $I_c(m)$ in the ensemble with a fixed number $m$ of erasures. Dotted gray line denotes the theoretical prediction, while colored dots are numerical results averaged over $>100$ samples at each $m$. (b) Coherent information $I_c(R \rangle Q M)$ with i.i.d. erasure errors, as a function of the erasure probability $p$. Gray line denotes the theoretical prediction for $N=13$, obtained from numerically integrating $I_c(pN)$ against a binomial distribution [see Eq.~\eqref{eq:Ic_erasure_sum}], while colored dots denote numerical results obtained by substituting the data in (a) into~\autoref{eq:Ic_erasure_sum}. (c) Eigenvalue density in the density matrix $\rho_K$, obtained from $\rho_Q$ by erasing $m$ qubits, for several values of $m$. Dashed lines denote the Marchenko-Pastur distribution with parameter $c = 2^{N - 1 - 2m}$, while solid lines denote numerical results aggregated across $>100$ samples at each $m$. The distribution is rescaled by the mean nonzero eigenvalue, as in~\autoref{fig:micro}a.}
    \label{fig:heralded}
\end{figure}

\autoref{fig:heralded}a compares our analytical prediction for $I_c(m)$ to numerical results in finite system sizes with $k = 1$ encoded qubits, showing excellent agreement. To obtain the full coherent information $I_c(R \rangle QM)$ for i.i.d. erasure channels, we must perform the sum in Eq.~\eqref{eq:Ic_erasure_sum}. For large but finite $N$, we expect the convolution with a binomial distribution to broaden the critical window from $N^{-1}$ to $N^{-1/2}$; this analytical prediction is similarly verified in \autoref{fig:heralded}b. In this figure, the ``theoretical'' curve denotes $I_c(m)$ numerically integrated over the distribution of $m$ values.

Finally, we briefly comment on the spectral statistics of the density matrix under erasure errors. Similar to the case of depolarization in the main text, a simple ansatz for the spectrum of $\rho_{K}$ for erasure numbers $m < m_c = \frac{1}{2}(N - k)$ is to regard $\rho_K$ as an incoherent mixture of $q^{m + k}$ random vectors. This suggests that the spectrum of $\rho_K$ is again distributed according to the Marchenko-Pastur distribution, with parameter $c = q^{N - k - 2m}$. Similarly, for $m > m_c$, we can instead regard $\rho_{RE}$ as an incoherent mixture of $q^{N - m}$ random vectors. Since the spectra of $\rho_K$ and $\rho_{RE}$ are identical, this suggests that the spectrum is distributed according to the Marchenko-Pastur distribution with parameter $c = q^{k + 2m - N}$. We verify this prediction for $k = 1$ encoded qubits in~\autoref{fig:heralded}c, where we see that the eigenvalue density is indeed quantitatively consistent with the Marchenko-Pastur distribution.

\color{black}
\section{Dual perspectives on the depolarizing channel}\label{app:dual}
The depolarizing channel can be expressed in two complementary ways, useful for examining different aspects of coding transitions.

Let 
\begin{equation}\label{eq:p-gamma}
\gamma \equiv \frac{q^2 }{q^2-1} p \in [0, 1].
\end{equation}
In terms of this parameter, the form of the local depolarizing channel presented in the main text becomes
\begin{equation}
    \mc{D}_{i ; \gamma}(\rho) \equiv \mathcal{N}_{i;p}(\rho) = \l 1 - \frac{\gamma(q^2-1)}{q^2} \r \rho + \frac{\gamma}{q^{2}}\sum_{\mu =1 }^{q^{2}-1} E^{\mu}_i\rho E^{\mu\dag}_i
    \label{eqn:dep_channel2-p}.
\end{equation} 
Expressing the channel in terms of Pauli errors at rate $p$ is natural for interpreting the ``band structure'' of $\tilde{\rho}_Q$ and $\tilde{\rho}_{RQ}$, as the analysis in the main text demonstrates. Borrowing language from statistical mechanics, we will refer to the standard depolarizing channel $\depol\l\rho\r = \bigotimes_{i \in Q} \mc{N}_{i; p }\l \rho \r$ as the \textit{canonical ensemble}, with $p$ (or $\gamma$) playing the role of inverse temperature. The fixed-weight error channel can then be viewed as a \textit{microcanonical ensemble}, with $w$ playing the role of energy. From this perspective, the ``temperature'' and ``energy'' are related through the equation $\overline{w} = p N$. 

An alternative perspective, useful for understanding dualities and higher-R\'enyi transitions, is that depolarization suppresses the amplitude of Pauli strings in the decomposition of $\rho$ (\autoref{eq:rho-pauli} below) according to their non-identity weight (cf.~\autoref{eq:depol-gamma}):
\begin{equation}
    \mathcal{D}_{i; \gamma} \l \rho\r = \l 1 - \gamma \r\rho + \gamma \, {\rm Tr}_{i}\lb \rho \rb\otimes \mathbbm{1}/q.
    \label{eqn:dep_channel}
\end{equation}

\section{Many-Body Weight Distributions and Weight Enumerator Polynomials}\label{app:mbw}
\subsection{Preliminaries}
Consider an arbitrary density matrix $\rho$ in a tensor product Hilbert space $\mathcal{H}_d$ composed of $d$ qudits. We can always decompose $\rho$ into a weighted sum of generalized Pauli strings: 
\begin{equation}\label{eq:rho-pauli}
    \rho = \frac{1}{q^d}\sum_{\stab\in \pg{d}}a_{\stab} \stab,
\end{equation}
where the $d$-qudit generalized Pauli group is defined using the same basis as the errors, as\footnote{We denote the Pauli operators that make up the state $\rho$ with the letter $\stab$, rather than $E$, to distinguish them from the errors that act upon this state, although of course this is the same basis.}
\begin{equation}
   \pg{d} = \lcb \stab_{\bm{\mu}} = \bigotimes_{i =1}^d E^{\mu_{i}} : \  \bm{\mu} \in \lb 0:q^2-1\rb^{d} \  \rcb.
\end{equation}
The normalization is chosen so that $a_I=1$. We can then take these strings and partition them into strings of fixed weight: let $\pg{d}_{w}$ be the set of generalized Pauli strings with Pauli weight $w$.

Associated to any quantum state we can then define a many body weight distribution (also called the operator size distribution~\cite{Schuster_23,Schuster_24,jacoby_24,zhang_24,Liu_2024}), $\phi$, as follows:
\begin{equation}\label{eq:purity}
    \phi\l w \r \equiv  \sum_{\stab \in \pg{d}_w} \lb a_{\stab}\rb^{2} .
\end{equation}
Using the identity $\tr{P P'} = q^d \delta_{P, P'}$, we can express the purity $\tr{\rho^2}$ in terms of the many-body weight $\phi(w)$ as follows:
\begin{equation}
    \tr{\rho^{2}} = \frac{1}{q^{2d}} \sum_{P, P' \in \pg{d}} a_P a_{P'} \tr{P P'} = \frac{1}{q^{2d}} \sum_{P \in \pg{d}} [a_P]^2 = \frac{1}{q^d} \sum_{w = 0}^d \phi(w) . 
\end{equation}
In other words, the purity is simply proportional to the total many-body weight. 

Since ${\rm Tr}_i[P]$ vanishes whenever $P$ has non-identity support on site $i$, the standard depolarizing channel induces a simple transformation on the many body weight distribution:
\begin{align}
\rho &\xrightarrow[]{\mc{D}} \frac{1}{q^d} \sum_{\stab \in \pg{d}} (1-\gamma)^{w(\stab)} a_\stab \stab \\
\label{eq:mbw-depol}
    \phi\l w \r &\xrightarrow[]{\mc{D}}  \l 1- \gamma\r^{2 w}  \phi \l w \r \equiv e^{-\eta w} \phi \l w \r, 
\end{align}
where $\eta = -2 \log{1 -\gamma } $, which is reminiscent of a thermodynamic identity. Note that in this way of decomposing the ``canonical ensemble,'' high and low temperature have been flipped: $(w,\eta)$ are now thermodynamically conjugate variables, so that increasing $\gamma$ \textit{decreases} the bias towards lower weights. We will return to this point below. 

\subsection{Application to quantum codes}
Consider a quantum code encoding a $k$ logical qudits. As illustrated in~\autoref{fig:noisy-encoding}, before acting with the quantum channel, this code is described by a pure state on the reference ($k$ qudits) + system ($N$ qudits). Writing each Pauli string in the decomposition of $\rho$ as a tensor product $\logic \otimes \stab$, we obtain
\begin{equation}\label{eq:pauli-decomposition}
\rho_Q = \frac{1}{q^N}\sum_{\stab \in \pg{N}}  a_{I,\stab} \stab,\qquad \rho_{RQ} = \frac{1}{q^{N+k}}\sum_{\logic \in \pg{k}} \sum_{\stab \in \pg{N}}  a_{\logic,\stab} \logic \otimes \stab.
\end{equation}
~\autoref{eq:pauli-decomposition} can be inverted to yield
\begin{equation}
    a_{I,\stab} = \tr{\stab \rho_Q}, \quad a_{\logic,\stab} = \tr{(\logic \otimes \stab) \rho_{RQ}}.
\end{equation}
Recalling~\autoref{eq:encoding-RQ}, for an arbitrary encoding unitary $U$, we further find
\begin{align}
q^k \tr{\stab \rho_Q \stab \rho_Q} &= q^k \sum_{n,m} \tr{(U^\dag \stab U) (\ket{n_L}\bra{m_L} \otimes \ket{0_A} \bra{0_A})} \tr{(U^\dag \stab U) (\ket{m_L}\bra{n_L} \otimes \ket{0_A}\bra{0_A})} \notag \\
&= \sum_{\logic \in \pg{k}} \tr{(U^\dag \stab U) (\logic \otimes \ket{0_A} \bra{0_A})}^2 = \sum_{\logic \in \pg{k}}\left[\tr{(\logic \otimes \stab) \rho_{RQ}}\right]^2 \notag \\
&= \sum_{\logic\in \pg{k}} [a_{\logic,\stab}]^2,
\end{align}
where $\ket{m_L},\ket{n_L}$ run over an orthonormal basis for the first $L$ qudits on the system\footnote{To go from the first line to the second line, we used the orthogonality property $\tr{P_1 P_2} = q^k \delta_{P_1 P_2}$ to transform from the operator basis $\ket{n}\bra{m}$ to the Pauli basis.}.

Stabilizer codes offer a familiar special case, in which $|a_{\logic,\stab}| \in \{0, 1\}$. In that case, $\logic$ labels the $q^{2k}$ logical cosets of the code: for each $\logic$, the set of $q^{N-k}$ operators $\stab$ for which $|a_{\logic,\stab}|=1$ are the operators that implement $\logic$ on the logical codespace.

Generalizing the many-body weight distribution introduced above, we can associate to any code --- stabilizer code or not --- a \textit{vector} of polynomials, $\phi_\logic(w)$, where 
\begin{equation}
\phi_{\logic}(w) = \sum_{\stab \in \pg{N}_w} a_{\logic,\stab}^2.
\end{equation}

The pair of \textit{weight enumerator polynomials} defined in Ref.~\cite{Shor_96}, commonly denoted $A(\cdot)$ and $B(\cdot)$, can be expressed as Laplace transforms of the $\phi_\logic(w)$:

\begin{subequations}
\begin{align}
A(u) &\equiv \sum_{\stab \in \pg{N}}   \tr{\stab \rho_Q} \tr{\stab^\dag \rho_Q} u^{w(\stab)} =\sum_{w = 0}^N \phi_I(w) u^w \label{eq:A}\\
B(u) &\equiv q^k \sum_{\stab \in \pg{N}}  \tr{\stab \rho_Q \stab^\dag \rho_Q} u^{w(P)} = q^k \sum_{w=0}^N \phi_{\Sigma}(w) u^w \label{eq:B}
\end{align}
\end{subequations}
where we have defined $\phi_{\Sigma}(w) = \sum_{\logic} \phi_{\logic}$. By our choice of normalization, $A(0) = B(0) = a_{I,I} = 1$, while $A(1)=q^{N-k}, B(1) = q^{N+k}$. For stabilizer codes, $A(u)$ enumerates the elements of the stabilizer group, while $B(u)$ enumerates the logical operators.

\subsection{Quantum MacWilliams Identity}

The dual interpretations of the depolarizing channel --- as suppressing the higher-weight Pauli strings in the decomposition of $\rho$ (\autoref{eq:mbw-depol}), or as applying Pauli errors with a probability given by the binomial distribution (\autoref{convex}) --- lead to the quantum MacWilliams identity as follows. Send $\rho_Q$ through a depolarizing channel, and evaluate the \textit{average fidelity}, defined as $\tr{\rho_Q \mathcal{D}_\gamma(\rho_Q)}$~\cite{Schumacher1996}.

By applying~\autoref{eq:mbw-depol},
\begin{equation}\label{eq:A-ave-fidelity}
    \tr{\rho_Q \mathcal{D}_\gamma (\rho_Q)} = q^{-N} \sum_\stab a_{I,\stab}^2 (1-\gamma)^{w(\stab)} = q^{-N} A(1-\gamma).
\end{equation}
On the other hand, applying~\autoref{convex},
\begin{align}\label{eq:B-ave-fidelity}
    \tr{\rho_Q \mathcal{N}_p(\rho_Q)} &= \sum_{E\in \pg{N}} \mathbbm{P}(E) \tr{E \rho_Q E^\dag \rho_Q} \notag \\
    &= q^{-k} (1-p)^N B\left(\frac{p}{(q^2-1)(1-p)}\right). 
\end{align}
Setting $u=\gamma/[q^2 - (q^2-1)\gamma]$ and using~\autoref{eq:p-gamma},
\begin{equation}\label{eq:macwilliams}
    A\left(\frac{1-u}{1+(q^2-1)u}\right) = q^{N-k} (1+(q^2-1)u)^{-N} B(u).
\end{equation}

Or, setting $u=1-\gamma$,
\begin{equation}\label{eq:macwilliams-2}
    B\left(\frac{1-u}{u(q^2-1)+1}\right) = q^{N+k}(1+(q^2-1)u)^{-N} A(u).
\end{equation}

This identity can be interpreted as a high-low temperature duality, relating the two perspectives on the depolarizing channel. Such dualities are common in statistical mechanics mappings for higher-R\'enyi entropies of decohered stabilizer codes, e.g., Ref.~\cite{FanBao_24,Su_24,Lyons_24,Niwa2025}. Indeed, the detection threshold and R\'enyi-2 threshold derived in the next section have both been addressed in the literature~\cite{FanBao_24,Su_24,Lyons_24,Turkeshi2024,Cao_24}. However, with the exception of Ref.~\cite{Cao_24}, those works make no mention of the quantum MacWilliams identity\footnote{The connection between Wegner's duality and the \textit{classical} MacWilliams identity was noted in Ref.~\cite{Kovalev2015}, but that identity applies to a special class of stabilizer codes, quantum CSS codes, discussed in the next subsection.}, instead appealing to more intricate statistical mechanics mappings or specific stabilizer models. We emphasize here that ~\autoref{eq:macwilliams} is a more general duality: it applies to any code defined as a projector onto some subspace, i.e. $\rho_Q = \Pi/q^k$, not just stabilizer codes. Its main limitation that, as $A(u)$ and $B(u)$ fundamentally depend on a second moment of the many-body weight distribution, the MacWilliams identity is only useful in probing transitions for certain R\'enyi indices ($\alpha = 2$ and $\alpha=\infty$), and does not constrain the true coding threshold.\footnote{One can express the logical failure probability or coherent information in terms of \textit{coset weight enumerators}, which however do not obey the MacWilliams identity~\cite{Cao_24}.}

\subsection{Aside: Classical MacWilliams identity and CSS codes}\label{app:classical}
The ability to write the error channel in two complementary ways is implicit in the construction of dual statistical mechanics models for decohered stabilizer codes. 
The most studied class of models in this context are (qubit) CSS codes~\cite{Calderbank1996,Steane1996}, whose stabilizer generators split into two sets, $X$-type and $Z$-type. The original statistical mechanics mapping works in the ``error-string picture'', mapping the probability of a logical class to the partition function of a spin model with quenched disorder set by a given error $E$. The optimal decoder then lives along the Nishimori line of this model, which for bit flip errors is $\exp(-2\beta) = \frac{p}{1-p}$~\cite{Dennis2002,Chubb_21,xu2025}, and the error correcting phase is the low-temperature, ordered phase. On the other hand, the ``stabilizer expansion'' model~\cite{FanBao_24,Lyons_24,Hauser2026,Su_24,Niwa2025} uses the classical analog of the many-body-weight distribution, resulting in Boltzmann factors $\exp(-2\beta) = 1-2p$. The latter model has the high-temperature paramagnetic phase as its coding phase. 

The relation $1-2p \leftrightarrow (1-p)/p$ is none other than the \textit{classical} MacWilliams identity~\cite{MacWilliams_book}. In CSS codes, it works in two ways: first, it acts as a high-low temperature duality between the error-string and stabilizer expansions for the state decohered by, say, $X$ errors. Second, since the two classical codes $\mathcal{C}_X$ and $\mathcal{C}_Z$ comprising a CSS code are each contained in the dual of the other, the classical MacWilliams identity is also a mapping between strong bit flip noise and weak phase flip noise~\cite{Kovalev2015}, valid only for certain R\'enyi indices~\cite{Su_24}.

\section{Consequences of the Quantum MacWilliams Identity}\label{app:macw}
The (zero-rate) detection threshold and the R\'enyi-2 threshold both follow from evaluating~\autoref{eq:macwilliams} at $u^*=1/(q+1)$:
\begin{equation}\label{eq:macw-crossing}
    A(u^*) = q^{-k} B(u^*).
\end{equation}

\subsection{Detection threshold}
As derived in Ref.~\cite{Ashikhmin,Scott_05}, the probability of a nondetectable logical error, averaged over pure code states $\ket{\psi}$, is
\begin{equation}
\mathbbm{P}_{nd}(p) = \mathbbm{E}_{\ket{\psi}} \left[\tr{(\Pi - \ket{\psi}\bra{\psi}) \depol(\ket{\psi}\bra{\psi})}\right] =  \frac{q^k (1-p)^N}{q^k + 1}\left[B\left(\frac{p}{(q^2-1)(1-p)}\right) - A\left(\frac{p}{(q^2-1)(1-p)}\right)\right].
\end{equation}
In the first expression, the average $\mathbbm{E}_{\ket{\psi}}$ is taken according to the Haar measure on the codespace defined by the projector $\Pi$; a few lines of algebra give rise to the prefactor $q^k/(q^k+1)$.

On the other hand, the probability to return to the codespace (zero syndrome) is proportional to the average fidelity introduced above:
\begin{equation}
\mathbbm{E}_{\ket{\psi}}\left[\tr{\Pi \depol(\ket{\psi}\bra{\psi})}\right] = q^k \tr{\rho_Q \depol(\rho_Q)} = (1-p)^N B\left(\frac{p}{(q^2-1)(1-p)}\right).
\end{equation}

Thus, the failure probability after postselecting on returning to the codespace is 
\begin{equation}\label{eq:fail-detect}
    P_{fail} =\frac{q^k}{q^k+1}\left(1 - \frac{A(u)}{B(u)}\right), 
\end{equation}
evaluated at $u=p/((q^2-1)(1-p))$.

~\autoref{eq:macw-crossing} then implies that at $p_d\equiv(q-1)/q$, the logical failure probability $P_{\rm fail}(p_d) = (q^k-1)/(q^k+1)$, independent of system size for zero-rate codes. In general, this marks a detection threshold, below which $P_{fail}$ is exponentially suppressed with system size, though for certain structured codes, the crossing lies within an intermediate, classical error-correcting phase. 

Further intuition for this threshold comes from specializing to stabilizer codes. In that setting, both $A(u)$ and $B(u)$ have a simple interpretation:
$$(1-p)^N A\left(\frac{p}{(q^2-1)(1-p)}\right)$$ is the total probability of a Pauli error $E$ belonging to the stabilizer group $\mathcal{S}$, and $$(1-p)^N B\left(\frac{p}{(q^2-1)(1-p)}\right)$$ is the total probability of an error $E\in \mathcal{N}(\mathcal{S})$. Thus, $1-A(u)/B(u)$ is the probability of a nontrivial logical operator being applied to the system, conditioned on measuring the all-zero syndrome~\cite{Cao_24}. Note that in deriving this threshold, we are using the weight enumerator polynomial to count the probabilities of the \textit{errors}. Logical information is lost when the probability of nontrivial logical operators becomes comparable to that of the trivial logical operators (i.e. stabilizers). 

The detection threshold of zero-rate codes also coincides with a threshold in the R\'enyi-$\infty$ coherent information~\cite{Su_24}
\begin{equation}\label{eq:Ic-infty}
    I_c^{(\infty)}(R\rangle Q) = -\glog_q[\lambda_{0}(\tilde{\rho}_Q)] + \glog_q[\lambda_{0}(\tilde{\rho}_{RQ})].
\end{equation}
Postselecting on returning to the codespace is not precisely equivalent to picking out the largest eigenvalue of the decohered density matrix, but it exhibits the same qualitative behavior, since at low $p$ the dominant contribution to the leading eigenvector comes from applying no error at all. Indeed, for Haar-random codes, we derived above using the Weingarten calculus that~\autoref{eq:Ic-infty} crosses zero at $H_\infty(p) = 1$, which yields $p_d = 1-1/q$. 

\subsection{Renyi-2 threshold}
On the other hand, we can use the weight enumerator polynomial to keep track of the suppression of large-size operators within $\rho$ (the other side of the duality). Taking $u=(1-\gamma)^2$, the ratio $A(u)/B(u)$ is now the ratio of purity-like quantities:
\begin{equation}
    \frac{A((1-\gamma)^2)}{B((1-\gamma)^2)} = \frac{\tr{\tilde{\rho}_0^2}}{\sum_{\bm{\mu}} \tr{\tilde{\rho}_{\bm{\mu}}^2}}
\end{equation}
where 
\begin{equation}
    \tilde{\rho}_{\bm{\mu}} = \mathrm{Tr}_R(E^{\bm{\mu}}_R \tilde{\rho}_{RQ}), \quad \bm{\mu} \in [0:q^2-1]^{\otimes k}
\end{equation}
From $u^* = 1/(q+1)$, this ratio has a crossing in zero-rate codes at $\gamma_2 = 1 - 1/\sqrt{q+1}$. 

To express this in more familiar terms, we compute the R\'enyi-2 entropies of $\tilde{\rho}_Q$ and $\tilde{\rho}_{RQ}$.
\begin{subequations}
\begin{align}\label{eq:renyi-AB}
S_2(\tilde{\rho}_Q) &= -\glog_q \tr{\tilde{\rho}_0^2} = N -\glog_q{A((1-\gamma)^2)}\\
S_2(\tilde{\rho}_{RQ}) &= k - \glog_q \left[\sum_{\vec{\mu}} \tr{\tilde{\rho}_{\vec{\mu}}^2 }\right] = N + k - \glog_q{B((1-\gamma)^2)}.
\end{align}
\end{subequations}

The R\'enyi-2 coherent information is therefore 
\begin{equation}\label{eq:renyi2-coherent}
I_c^{(2)}(R\rangle Q) =  \mathrm{log}_q\left[\frac{B((1-\gamma)^2)}{A((1-\gamma)^2)}\right] - k.
\end{equation}
which is precisely zero at $\gamma = \gamma_2$. Note that for qubits, $p_2 = (3-\sqrt{3})/4$ lies along the self-dual surface obtained in Ref.~\cite{Su_24} for stabilizer codes under generic Pauli noise; the entire self-dual surface can be obtained by appealing to yet another quantum MacWilliams identity, for a pair of \textit{complete weight enumerator polynomials}~\cite{Du_23}. This result is also consistent with the argument from the band structure in the main text, as well as the derivation from Weingarten calculus in~\autoref{app:weingarten} of this Supplement, upon noting that $H_2((q^2-1)\gamma_2/q^2) = 1$.

The fact that both the R\'enyi-2 and R\'enyi-$\infty$ thresholds are derived from a single identity reflects a close connection between the two quantities, which previously has been derived for stabilizer codes~\cite{Su_24}. In particular, since both quantities depend on the ratio $A(u)/B(u)$, the scaling behavior of this ratio near $u=u^*$ imbues the two thresholds with the \textit{same} universal critical behavior.

\section{Weight Enumerators in Haar-random codes}\label{app:rmt}
While the quantum MacWilliams identity fixes the location of both the detection threshold and the R\'enyi-2 threshold, it tells us nothing about the scaling behavior near the threshold. In this section, we derive an exact expression for the Haar average of $A(u)$ and $B(u)$ and compare it to the ansatz for the spectrum obtained in the main text, with which it agrees to leading order. 

\subsection{Haar average of $A(u)$ and $B(u)$}
To determine $A(u)$ in a Haar-random code, we take the average both sides of~\autoref{eq:A-ave-fidelity}:
\begin{equation}
\haar \tr{\rho_Q \mathcal{D}_\gamma(\rho_Q)} = q^{-N} \haar A(1-\gamma).
\end{equation}
The left-hand side is the Haar average of a two-replica quantity, so a short calculation in the spirit of~\autoref{app:weingarten} of this Supplement yields
\begin{equation}\label{eq:A-haar}
\haar A(1-\gamma) = \frac{q^{N-k}}{q^{2N}-1} \left(q^N(1-p)^N (q^N - q^k) + q^{N+k}-1\right).
\end{equation}
In this subsection, to highlight the connections to the many-body weight distribution, we present an alternative method for determining (the Haar average of) $A(u)$, in the form of a random-matrix-theory-inspired ansatz.\footnote{As $A(u)$ is self-averaging, we will henceforth drop the notation $\haar$.}

From~\autoref{eq:A}, the functional form of $A(u)$ is uniquely determined by $\phi_I(w)$. The many-body weight distribution of a random Hermitian matrix (drawn from the Gaussian unitary ensemble) on $d$ qudits is a Gaussian peaked at $\overline{w} = \frac{q^2-1}{q^2}$. The density matrix $\rho_{RQ}$ is not quite a random Hermitian matrix, though: it is a positive semi-definite Hermitian matrix with trace 1. 

We therefore obtain a minimal ansatz for $\phi_I(w)$ by imposing two constraints. First, 
\begin{equation}
\tr{\rho_Q} = 1 \Rightarrow \phi_I(0) = 1.
\end{equation}
Second, from~\autoref{eq:purity},
\begin{equation}\label{eq:purity-constraint}
\tr{\rho_Q^2} = q^{-k} \Rightarrow \sum_w \phi_I(w) = q^{N-k}.
\end{equation}
To satisfy both constraints while keeping the many-body weight distribution close to binomial, we take
\begin{equation}\label{eq:phi-ansatz}
\phi_I(w) = \begin{cases}
1 & w=0 \\
a \binom{N}{w} (q^2-1)^w  & w>0
\end{cases}.
\end{equation}
The constant $a$ is fixed by substituting into~\autoref{eq:purity-constraint}, yielding $a=\frac{q^{N-k}-1}{q^{2N}-1}$.

Substituting into the definition of the A-type polynomial (\autoref{eq:A}), and applying the MacWilliams identity~\autoref{eq:macwilliams-2} yields
\begin{equation}\label{eq:AB-ansatz}
A(u) = 1-a + a(1 + (q^2-1)u))^N, \qquad B(u) = q^{N+k} a + q^{k-N} (1-a) (1 + (q^2-1)u)^N,
\end{equation}
in agreement with~\autoref{eq:A-haar}. This should come as no surprise: since $A(u)$ is related to a second moment over the Haar distribution, it suffices to impose two constraints. Since the Clifford group is a two-design,~\autoref{eq:AB-ansatz} also applies to random stabilizer codes. 

Substituting~\autoref{eq:AB-ansatz} into~\autoref{eq:renyi-AB} yields the theoretical curve for $S_2(\tilde{\rho}_Q)$ in~\autoref{fig:postselect-renyi}(c) of the main text. We observe excellent agreement between the theoretical curve, which is technically an annealed average, and the numerical data, which is a quenched average over 200 samples. 

Under the numerically supported assumption that $A(u)$ and $B(u)$ are self-averaging, we can read off the critical behavior of both the R\'enyi-2 threshold and the detection threshold for zero-rate codes from~\autoref{eq:AB-ansatz}. Expanding $A(u)/B(u)$ in the vicinity of $u^*$,
\begin{equation}
\frac{A(u)}{B(u)} = q^{-k} - (u-u^*)N \left[\frac{(q^2-1)(q^k-1)(q^N+1)}{q^{k+1}(q^k+1)(q^N-1)}\right] + O((u-u^*)^2).
\end{equation}
For $k=O(1),N\rightarrow\infty$, the term in brackets approaches an $O(1)$ number, such that the scaling function to leading order is $f((u-u^*)N)$. This is consistent with the ``$\nu=1$'' scaling collapse of the $\alpha=2$-reweighted coherent information presented in the main text [\autoref{fig:postselect-renyi}(d)]. 

\subsection{Perturbation theory ansatz, revisited}
With~\autoref{eq:AB-ansatz} in hand, we can now compare it to the perturbation theory ansatz in the main text and~\autoref{app:pert_ansatz} of this Supplement.

We focus on the leading eigenvalues of $\tilde{\rho}_{RQ}$ and $\tilde{\rho}_{Q}$, respectively. Plugging $w=0$ into~\autoref{eqn:msa} and~\autoref{eq:msa-Q}, we obtain
\begin{subequations}
\begin{align}
    \lambda_{RQ}^{(0)} &= (1-p)^N + \sum_{w'>0} \frac{P_{w'}}{q^{N+k}} = (1-p)^N \left[1 - \frac{1}{q^{N+k}}\right] + \frac{1}{q^{N+k}}  \\
    \lambda_{Q}^{(0)} &= \frac{(1-p)^N}{q^k} + \sum_{w'>0} \frac{P_{w'}}{q^N} = \frac{1}{q^k}\left[(1-p)^N \left(1 - \frac{1}{q^{N-k}}\right) + \frac{1}{q^{N-k}}\right].
\end{align}
\end{subequations}
On the other hand, evaluating $A(u)$ and $B(u)$ evaluated at $u=p/((q^2-1)(1-p))$ yields
\begin{subequations}
\begin{align}
    (1-p)^N A\left(\frac{p}{(q^2-1)(1-p)}\right) &= (1-p)^N \left[1 - \frac{1}{q^{N+k}} \textcolor{red}{\frac{1-q^{k-N}}{1-q^{-2N}}}\right] + \frac{1}{q^{N+k}}\textcolor{red}{\frac{1-q^{k-N}}{1-q^{-2N}}} \\
    (1-p)^N B\left(\frac{p}{(q^2-1)(1-p)}\right) &= (1-p)^N \left(1 - \frac{1}{q^{N-k}}\right) \textcolor{red}{\frac{1}{1 - q^{-2N}}} + \frac{1}{q^{N+k}} \left[1 - \textcolor{red}{\frac{1}{q^{N+k}} {\frac{1-q^{k-N}}{1-q^{-2N}}}}\right].
\end{align}
\end{subequations}
Thus, up to subleading order in $q^{-N}$, $(1-p)^N A(u)$ matches our ansatz for the leading order eigenvalue of $\tilde{\rho}_{RQ}$, while $q^k (1-p)^N B(u)$ matches the ansatz for the (approximately $q^k$-fold degenerate) leading eigenvalue of $\tilde{\rho}_Q$. Each term colored red is a correction to the ansatz.

For finite-rate codes, the crossing fixed by the MacWilliams identity at $p=1-1/q$ is not especially meaningful, as the postselected failure probability is already exponentially small at $p=1-1/q$. A more sensible definition of the detection threshold is as the point  where (in the limit $k,N\rightarrow\infty, k/N=r$), $P_{fail}$ changes from $O(1)$ to exponentially small, or equivalently, as where $S_\infty(\tilde{\rho}_Q)$ has a non-analyticity: While that threshold is not fixed by the quantum MacWilliams identity, it can be identified within the RMT ansatz,
\begin{equation}
H_\infty(p_d(r)) = \glog_q(1-p_d) = 1 - r \Rightarrow  p_d(r) = 1-q^{r-1}.
\end{equation}
This rate-dependent threshold was obtained previously in Ref.~\cite{Turkeshi2024}, and indeed, plugging~\autoref{eq:AB-ansatz} into~\autoref{eq:fail-detect}, we recover the annealed average for the postselected fidelity obtained in that work\footnote{That is, $P_{fail} = 1-\tilde{F}$, see Eq. 8 of Ref.~\cite{Turkeshi2024}.}.

\color{black}
\section{Band structure in stabilizer codes}\label{app:stabilizer}

Our focus in this work has been on the band structure of unstructured, Haar-random codes. A natural question is to what extent a similar framework applies in more structured settings, e.g., LDPC stabilizer codes. For simplicity, we will set $q=2$. 

\subsection{Spectrum of $\tilde{\rho}_Q$}
An $[[N,k,d]]$ stabilizer code is characterized by a stabilizer group $\mathcal{S}$, generated by $N-k$ independent commuting Pauli operators~\cite{Gottesman2024}
\begin{equation}
\mathcal{S} \equiv \langle g_1,g_2,...,g_{N-k} \rangle, \qquad g_i^2 = \mathbbm{1}, \quad [g_i,g_j] = 0.
\end{equation}
The distance $d$ is the minimum weight of a Pauli operator that commutes with all of $\mathcal{S}$ but does not belong to $\mathcal{S}$. A stabilizer code is said to be \textit{non-degenerate} if no element of $\mathcal{S}$ (besides the identity) has weight $< d$. The term non-degenerate can also be applied to a code (stabilizer or otherwise) with reference to a particular error set $\mathcal{E}$, if distinct errors in that set map distinct codewords onto orthogonal states (cf.~\autoref{app:kl} of this Supplement).\footnote{For a more thorough introduction to stabilizer codes, presented in a similar language to that used here, we refer the reader to Chapter 3 of Ref.~\cite{Gottesman2024}.}

The code space is the joint +1 eigenspace of $\mathcal{S}$; we will denote the projector onto this code space $\Pi_{\bm 0}$, so that
\begin{equation}
    \rho_Q = \frac{1}{2^N} \prod_{i=1}^{N-k} (\mathbbm{1} + g_i) = \frac{\Pi_{\bm{0}}}{2^k}.
\end{equation}
Note that expanding out the product yields the Pauli string decomposition in~\autoref{eq:pauli-decomposition}, with $a_{I,P}=1$ if and only if $P \in \mathcal{S}$.

Now suppose we act with a Pauli error $E$. The syndrome $\bm{s}$ of this error is the set of stabilizer generators that anticommute with it, expressed as a length $(N-k)$ bit vector. This error takes us to the subspace defined by $\Pi_{\bm s}$:
\begin{equation}
    \Pi_{\bm s} = \prod_{i=1}^{N-k} \frac{\mathbbm{1} + (-1)^{s_i}g_i}{2}.
\end{equation}

Therefore, acting with the full depolarizing channel $\depol$, the resulting mixed state $\tilde{\rho}_Q$ decomposes into a convex sum of the orthogonal projectors $\Pi_{\bm s}$, weighted by the syndrome probabilities $\mathbbm{P}(\bm s)$:
\begin{equation}\label{eq:rhoQ-stab}
\tilde{\rho}_Q = \frac{1}{2^k} \sum_{\vec s} \mathbbm{P}(\bm s) \Pi_{\vec{s}}.
\end{equation}
Accordingly, the spectrum of $\tilde{\rho}_Q$ splits into $2^{N-k}$ sets of $2^k$ exactly degenerate eigenvalues, one set for each syndrome: $\lambda_{\vec{s}} = \mathbbm{P}(\bm s)/2^k$. 

In non-degenerate codes at sufficiently low $p$, the syndrome probability is dominated by the lowest-weight error with that syndrome. Thus, the spectrum organizes into ``syndrome bands'' where band $w$ contains all the syndromes with minimum-weight-error $w$: precisely the zeroth-order ansatz for the spectrum of a decohered Haar-random code. As $p$ increases, the contributions from higher-weight errors become non-negligible, with two effects: (1) the eigenvalues are shifted, (2) the bands generically acquire a dispersion, as different syndromes with the same minimum-weight error may have different multiplicities of higher-weight errors. 

\subsection{Spectrum of $\tilde{\rho}_{RQ}$}
We now consider $\tilde{\rho}_{RQ}$, whose spectrum is governed not only by the syndrome probabilities, but by the probabilities of different logical classes conditioned on the syndrome. 

To unpack that statement, recall that logical operators are Pauli operators that commute with $\mathcal{S}$, thus leaving the state within the codespace. Such an operator $L$ either acts trivially on the codespace (if $L \in \mathcal{S}$), or applies a nontrivial logical transformation. The $2^k$-dimensional logical group is therefore the quotient group $\mathcal{N}(\mathcal{S})/\mathcal{S}$, where $\mathcal{N}(G)$ is the normalizer of group $G$. 

We now define the logical operators $(\overline{X}_i, \overline{Z}_i), i=1,...,k$, which implement logical $X_i, Z_i$ operations on the code space. These obey the anticommutation relations $[[\overline{X}_i, \overline{Z}_j]] = (-1)^{\delta_{ij}}$, where the scalar commutator $[[A,B]]$ is defined by the equation $AB = [[A,B]] BA$~\cite{Chubb_21}. This choice of logical representatives yields, for the pure state $\rho_{RQ}$
\begin{equation}
    \rho_{RQ} = \frac{1}{2^{N+k}} \prod_{i=1}^{k} (\mathbbm{1}_{RQ} + Z_i \otimes \overline{Z}_i)(\mathbbm{1}_{RQ} + X_i\otimes \overline{X}_i) \prod_{i=1}^{N-k}(\mathbbm{1}_{RQ} + \mathbbm{1}\otimes g_i),
\end{equation}
where each tensor product is on the space $R\otimes Q$. Note that expanding out the product yields the Pauli string decomposition in~\autoref{eq:pauli-decomposition}, with $a_{P',P}=1$ if and only if $P$ is stabilizer-equivalent to the canonical logical operator $\overline{P'}$.

Given a syndrome $\vec{s}$, the decoder's objective is to apply a correction operator that returns to the state to the codespace (i.e., an operator with syndrome $\vec{s}$) \textit{without} implementing a nontrivial logical operation. After suffering an error $E$, the syndrome $\vec{s}$ divides into $4^k$ logical classes, or cosets, differentiated by their ``logical syndrome'' (not known to the decoder) $\vec{l} = ([[\overline{X}_1, E]], [[\overline{Z}_1, E]],...)$. A maximum likelihood decoder works by computing the relative probabilities of each of these $4^k$ logical classes, and choosing a correction operator in the most likely logical class. Decoding is successful if the actual error that occurred belongs to this most likely class. Conditioned on syndrome $\vec{s}$, the recovery probability is therefore
\begin{equation}\label{eq:Psucc-s}
    P_{\rm success}(\bm s) = \frac{\max_{\bm l} \mathbbm{P}(\bm s, \bm{l})}{\mathbbm{P}(\bm{s})} = \max_{\bm l} \mathbbm{P}(\bm{s}|\bm{l}),
\end{equation}
and the average recovery probability
\begin{equation}\label{eq:Psucc-ave}
\langle P_{\rm success}(\vec{s})\rangle_{\vec{s}} = \langle \max_{\bm l} \mathbbm{P}(\bm{s}|\bm{l})\rangle_{\vec{s}} = \sum_{\vec{s}} \max_l \mathbbm{P}(\bm{s}, \bm{l}),
\end{equation}
where $\langle \cdot\rangle_{\vec{s}}$ denotes the average over syndromes, weighted by their respective probabilities.

We can relate this distribution of joint syndrome and logical class probabilities to the spectrum of the decohered state on the joint system and reference as follows. Let $\overline{P}^{(\bm l)}$ denote a logical operator with logical syndrome $\vec{l}$, which acts as $P^{(l)}$ on the logical degrees of freedom. Also let $E^{(\bm s)}$ be a ``canonical'' error with syndrome $\vec{s}$, that is, any one of the $2^{N-k}$ stabilizer-equivalent Pauli operators with syndrome $\vec{s}$ and logical syndrome $\vec{0}$. Defining
\begin{equation}
    \ket{\bm{s},\bm{l}} = P^{(\bm l)}_R \otimes \left(\overline{P}^{(\bm l)}E^{(\bm s)}\right)_Q \ket{\Psi}_{RQ}
\end{equation}
we obtain, for the decohered state on the system + reference\footnote{See Ref.~\cite{Colmanarez2024,Niwa2025,Colmanarez2025} for similar discussions, with the latter two works specializing to CSS codes.}
\begin{equation}\label{eq:rhoRQ-stab}
\tilde{\rho}_{RQ} = \sum_{\vec{s},\vec{l}} \mathbbm{P}(\bm{s}, \bm{l}) \ket{\bm{s},\bm{l}} \bra{\bm{s},\bm{l}}.
\end{equation}
\autoref{eq:rhoRQ-stab} implies that, for each set of $2^k$ degenerate eigenvalues of $\tilde{\rho}_Q$ corresponding to a syndrome $\vec{s}$, there are associated $4^k$ eigenvalues of $\tilde{\rho}_{RQ}$, which are precisely the joint probabilities $\mathbbm{P}(\bm{s},\bm{l})$. \autoref{eq:Psucc-s} says that a given syndrome can be accurately decoded with high probability if one of these $4^k$ eigenvalues is much larger than the others, while \autoref{eq:Psucc-ave} says that the code is approximately correctable if this separation of eigenvalues holds for \textit{typical} syndromes, i.e., those that dominate the spectrum of $\tilde{\rho}_Q$.

\subsection{Examples}
\begin{figure}[t]
\includegraphics[width=\linewidth]{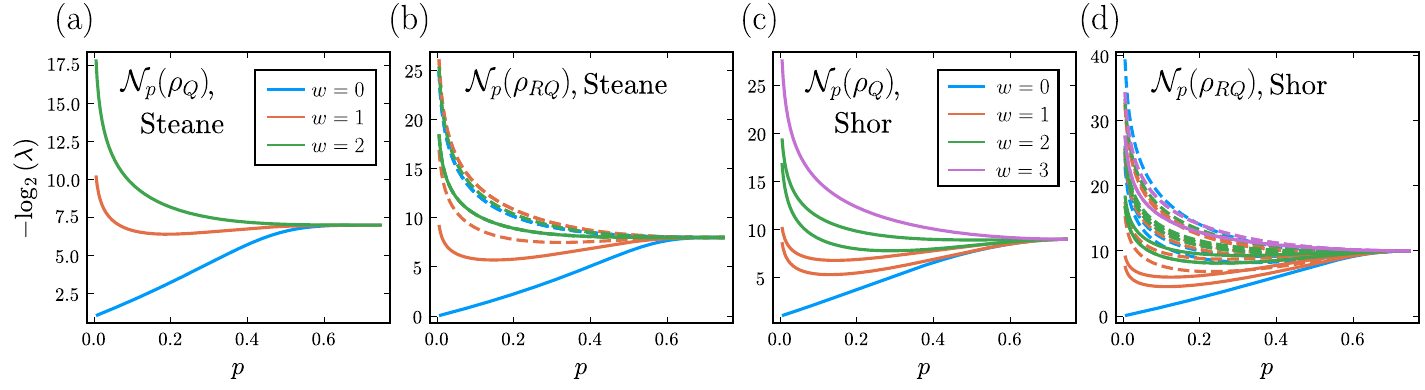}
\caption{\label{fig:shor-steane} Spectra of the [[7,1,3]] Steane and [[9,1,3]] Shor codes subject to depolarizing noise at rate $p$. (a) Eigenvalues of $\mathcal{N}_p(\rho_Q)$ for the Steane code fall into 3 exactly degenerate bands, associated with weight 0, 1, and 2 errors. (b) Eigenvalues of $\mathcal{N}_p(\rho_{RQ})$ for the Steane code. Each syndrome leads to 4 eigenvectors in $\mathcal{N}_p({\rho}_{RQ})$; the leading eigenvalue is shown as a solid curve, while the remaining eigenvalues are shown as dashed curves. For syndromes in the $w=2$ band, the leading eigenvalue is three-fold degenerate. (c), (d) Same as (a), (b) but for the Shor code. The $w=1$ and $w=2$ ``bands'' split according to the degeneracy of the multiplicity of degenerate errors.}
\end{figure}
The above phenomenology is best illustrated by two minimal examples. We first consider the $[[7,1,3]]$ Steane code, which is non-degenerate with respect to single-qubit errors. There are $2^{N-k}$ different syndromes, which fall into 3 different bands in $\tilde{\rho}_Q$: (i) the $w=0$ band, with degeneracy 2, associated with the syndrome $\vec{s}=\vec{0}$, (ii) the $w=1$ band, with degeneracy $42$, associated with the $21$ syndromes arising (predominantly) from single-qubit errors, (iii) the reservoir, containing the 84 remaining eigenvalues, all degenerate, and all dominated by $w=2$ errors. 

This spectrum, shown as a function of $p$ in \autoref{fig:shor-steane}a, has two salient features. First, each band is ``flat'' in the sense that the eigenvalues are completely degenerate: due to the symmetries of the code, all $w=1$ syndromes have the same probability, and all the remaining nontrivial syndromes are likewise equally probable. That is, two syndromes with the same minimal error weight ($w=0, w=1, w=2$) also have precisely the same composition of higher-weight errors.

Turning to $\tilde{\rho}_{RQ}$ (\autoref{fig:shor-steane}b), the correctability of the $w=0$ and $w=1$ bands at sufficiently low $p$ is evidenced by the fact that for each such syndrome, the leading eigenvalue (solid curves) is significantly larger than the remaining $2^k-1=3$ eigenvalues (dashed curves). Thus, the leading eigenvalues of the correctable bands form well-separated bands, while the subleading eigenvalues, along with all the eigenvalues of the partially filled, uncorrectable $w=2$ band, belong to the reservoir.

A slightly richer example is the $[[9,1,3]]$ Shor code, which is degenerate with respect to phase flip errors. Degeneracy leads to a splitting of the nominal $w=1$ and $w=2$ bands in \autoref{fig:shor-steane}c. The $w=1$ band is split because single-qubit errors come in two types: single-qubit $X$ and $Y$ errors are distinguishable, leading to 18 distinct syndromes and 36 eigenvalues in the higher orange curve ($\lambda\sim p/6$); whereas single-qubit $Z$ errors are correctable but only map to 3 distinct syndromes, leading to 6 eigenvalues in the lower orange curve ($\lambda \sim p/2$). A similar phenomenon occurs at $w=2$, which splits into $\lambda \sim p^3/18$ and $\lambda \sim p^2/3$ respectively. 

As with the Steane code, the leading eigenvalues of the $w=0$ and $w=1$ bands of $\tilde{\rho}_Q$ map onto well-defined bands of $\tilde{\rho}_{RQ}$, with the subleading eigenvalues belonging to the reservoir. The fate of the $w=2$ band is less discernible. For this simple example, it turns out that the less degenerate $w=2$ syndromes (those with probability $\sim 3 p^2$) are correctable at low $p$ (have a distinguished leading eigenvalue), while the \textit{more probable} $w=2$ syndromes are not (the leading eigenvalue is degenerate, meaning two of the four logical classes are equally likely). 

This minimal example of degeneracy provides some insights and caveats for the application of band structure to families of degenerate codes. First, we see that band structure based on the lowest-weight error contributing to that syndrome can survive some degeneracy, so long as the degeneracy is small enough: roughly, if the number of minimal-weight errors contributing to a syndrome is less than $3(1-p)/p$, then the band splitting will be less than the gap to the next band. Of course, for a growing family of degenerate codes, this will break down if the degeneracy scales with $N$.

Second, the Shor code provides an example where $\alpha-$reweighting the spectrum of $\tilde{\rho}_Q$ may have an adverse effect: the $w=2$ band contains some correctable syndromes, but these have \textit{lower} eigenvalues than the uncorrectable $w=2$ syndromes. This suggests that the R\'enyi threshold can be made non-monotonic in $\alpha$ by injecting a large enough density of ``bad'' errors into high-probability bands.